\documentclass[12pt,a4paper]{article}
\pdfoutput=1
\usepackage{jheppub}
\usepackage[utf8]{inputenc}
\usepackage[T1]{fontenc}
\usepackage{comment}

\usepackage{amsmath,physics,float}  
\usepackage{amssymb}  
\usepackage{latexsym}
\usepackage{graphicx,xcolor}

\usepackage{amsfonts}
\usepackage{braket}
\usepackage{mathrsfs}

\usepackage{subfig} 
\usepackage{empheq}

\usepackage[shortlabels]{enumitem}

\usepackage{pgfplots}
\usepackage{natbib}

\usepackage{bigints}

\newcommand{\beq}{\begin{equation}}
\newcommand{\eeq}{\end{equation}}

\newcommand{\lc}{\left(}
\newcommand{\rc}{\right)}
\newcommand{\ls}{\left[}
\newcommand{\rs}{\right]}

\usepackage{tikz, pgf}
\usetikzlibrary{decorations.pathmorphing}
\usetikzlibrary{arrows.meta}
\tikzset{%
  >={Latex[width=2mm,length=2mm]},
            base/.style = {rectangle, rounded corners, draw=black,
                           minimum width=4cm, minimum height=1cm,
                           text centered, font=\sffamily},
}
\usetikzlibrary{calc}
\begin{document}

\title{\boldmath Critical exponents for higher order phase transitions: Landau theory and RG flow}

\author[1]{Joydeep Chakravarty,}
\emailAdd{joydeep.chakravarty@icts.res.in}

\author[2]{Diksha Jain}
\emailAdd{diksha.2012jain@gmail.com}

\affiliation[1]{International Centre for Theoretical Sciences (ICTS-TIFR)\\
Tata Institute of Fundamental Research\\
Shivakote, Hesaraghatta,
Bangalore 560089, India.}
\vspace{4pt}

\affiliation[2]{International Centre for Theoretical Physics\\
Strada Costiera 11, Trieste 34151 Italy
\vspace{4pt}}

\begin{abstract}
{In this work, we define and calculate critical exponents associated with higher order thermodynamic phase transitions. Such phase transitions can be classified into two classes: with or without a \textit{local} order parameter. For phase transitions involving a local order parameter, we write down the Landau theory and calculate critical exponents using the saddle point approximation. Further, we investigate fluctuations about the saddle point and demarcate when such fluctuations dominate over saddle point calculations by introducing the generalized Ginzburg criteria. We use Wilsonian RG to derive scaling forms for observables near criticality and obtain scaling relations between the critical exponents. Afterwards, we find out fixed points of the RG flow using the one-loop beta function and calculate critical exponents about the fixed points for third and fourth order phase transitions. }
\end{abstract} 

\maketitle
\pagebreak

\section{Introduction}
A thermodynamic phase transition is characterized by the non-analyticity of thermodynamic observables with respect to temperature across two or more different phases. These phase transitions were systematically classified by Ehrenfest \cite{Ehrenfest, articlegj}. Ehrenfest's classification is based on the \textit{order} of the phase transition, i.e. a $r$th order phase transition is characterized by the $r$th derivative of the partition function (or free energy) becoming discontinuous at the temperature where the phase transition takes place. A thermodynamic function that takes different values across different phases is called an order parameter.

With regards to second order phase transitions, Landau put forward a phenomenological Hamiltonian which captures essential aspects of the transition in the infrared limit of the system \cite{Landau:1937obd}. His formalism suggests that near the critical temperature $T_C$ (the temperature at which phase transition takes place), the Hamiltonian can be expanded in terms of a \textit{local} order parameter $\phi (x)$ on the basis of appropriate symmetries. For example, if a system has a $Z_2$ symmetry i.e. it is symmetric under $\phi(x) \to -\phi(x)$; then for $T \approx T_C$ and a small external coupling field $h$, the Hamiltonian $H(\phi)$ can be expressed as follows:

\beq
\beta H =\int d^dx\lc \frac{K}{2}(\grad \phi)^2 + t \phi^2(x) + u \phi^4(x) + \dots - h.\phi(x) \rc
\eeq
where $t$ and $u$ are couplings which are functions of the temperature. Here we have restricted ourselves to only the lowest order terms. It can be readily demonstrated that the partition function computed using this Hamiltonian gives rise to different values of the order parameter for $t<0$ and $t>0$ using the saddle point approximation. The divergences of observables near the critical point are captured using \textit{critical exponents}.

To obtain the correct picture of fluctuations away from the saddle point, we need to incorporate the field-theoretic machinery of the renormalization group (RG) flows of couplings in the theory. We can use the behavior of couplings under RG flow to determine how thermodynamic observables scale in the vicinity of the critical surface. Consequently, we can derive various scaling relations between the theory's critical exponents, which are robust under fluctuations away from the saddle point.

In order to understand the RG flows in the coupling space, we start by finding out fixed points of the flow and then expand the couplings about the fixed points. This leads to a classification of the couplings into relevant, marginal, and irrelevant depending on their behavior in the infrared (IR). In the IR, relevant couplings diverge away from the IR fixed points as we integrate out degrees of freedom while irrelevant couplings converge towards the IR fixed point. The subspace of relevant couplings is usually finite-dimensional, and hence theories in the IR are characterized by only a finite number of relevant couplings. Such IR behaviour defines the notion of \textit{universality classes} \cite{Hollowood:2009eh, Peskin:1995ev, kardarcourse, kardar_2007}. The set of relevant couplings can be used to determine some of the critical exponents, and scaling relations between them further allows us to determine all the critical exponents \cite{doi:10.1063/1.1734338, JOSEPHSON1966608,doi:10.1063/1.1696618, kardarcourse, kardar_2007}.

\begin{center}

\begin{figure}
\begin{tikzpicture}[node distance= 2cm,
    every node/.style={fill=white, font=\sffamily}, align=center]
  \node (start)             [base]              {\textbf{Higher order Phase transitions}};
  \node (lop)     [base, below of=start]          {\textit{Local order parameter}};
  \node (nlop)     [base,  xshift= 6cm]          {\textit{Non-local order parameter:}\\
  (To be discussed in\\
  forthcoming work)};
  \node (dimension1)     [base, below of=lop, xshift= -4.5cm]          {$d > 2r$};
  \node (dimension2)     [base, below of=lop, xshift= 4.5cm]          {$d \leq 2r$};
  
  \node (saddle1)      [base, below of=dimension1,    yshift= -2.5cm]   {Saddle Point approximation \\valid, critical exponents\\  given in Table \ref{table1}};
   \node (saddle2)      [base, below of=dimension2, xshift= -2.5cm,  yshift= -2.5cm]   {Saddle Point valid,\\ critical exponents are \\ given in Table \ref{table1}};
   \node (saddle3)      [base, below of=dimension2, xshift= 3cm,  yshift= -2.5cm]   {Fluctuations dominate over saddle\\ point, Scaling relations found \\ using RG flow in \S\ref{secscaling} (Table \ref{table2}),\\
   Critical exponents given in Table \ref{table5}};
  
  \draw[->]             (start) -- (lop);
  \draw[->]             (start) -- (nlop);
  \draw[->]     (lop) -| (dimension1);
\draw[->]     (lop) -| (dimension2);
\draw[->]     (dimension1)-- (saddle1);
\draw[->]     (dimension2)-| node[yshift = -1.5cm]{Follows \\ Ginzburg Criteria (\S\ref{Ginzburg})} (saddle2);
\draw[->]     (dimension2)-|  node[yshift = -1.5cm]{Does not follow \\Ginzburg criteria} (saddle3);  \end{tikzpicture}
\caption{A flowchart describing the general outline of the paper.} 
\label{flowchart}
\end{figure}
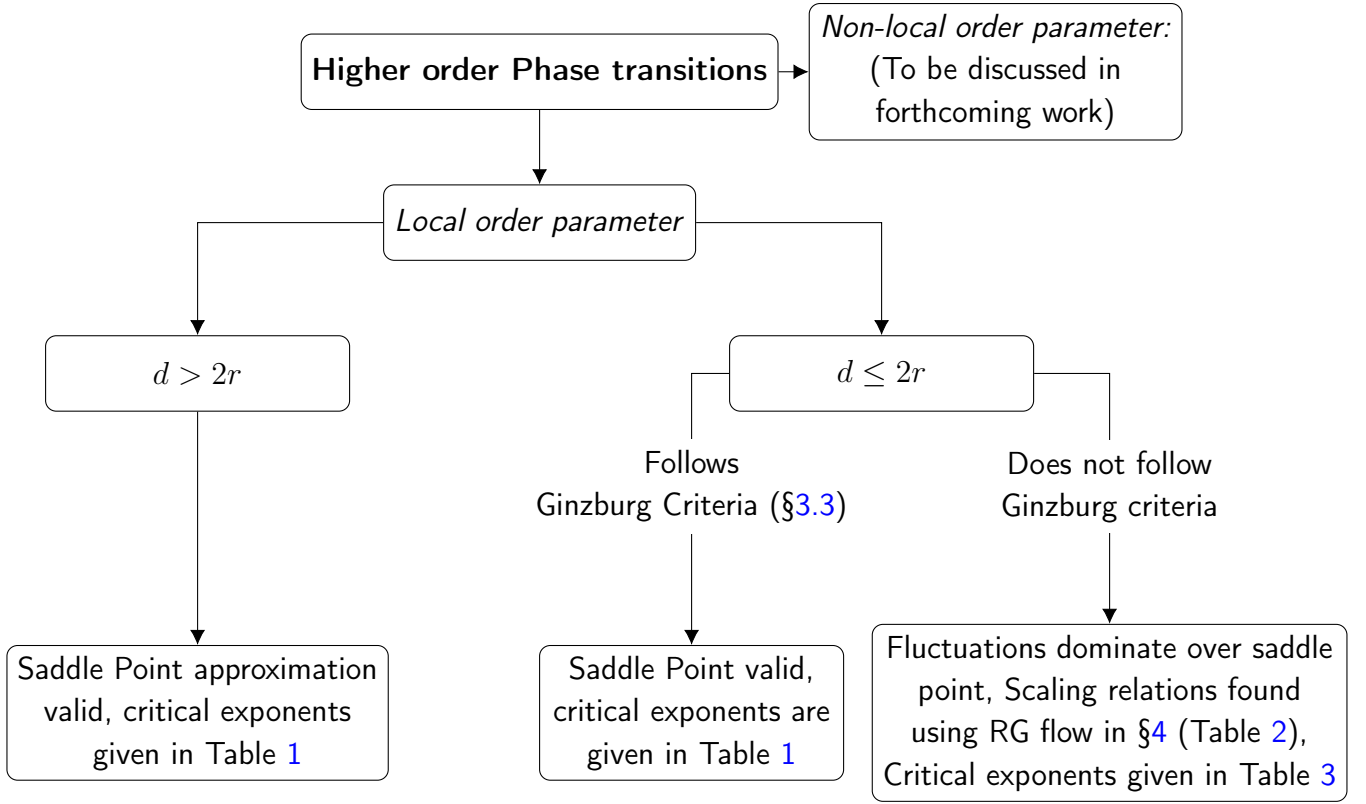
    
\end{center}

We will now briefly describe our work, a summary of which is given in Fig. \ref{flowchart}. In the first part of our work, we apply various aspects of our above discussion to the study of higher order phase transitions described by local order parameters. In \S \ref{Landauth} we generalize Landau's phenomenological Hamiltonian for higher-order phase transitions using a local order parameter such that physical observables derived from the partition function have analytic properties in the $h-t$ plane apart from a branch cut ending at the critical point. Next, we define critical exponents for higher order phase transitions in \S \ref{seccrit} and calculate them using saddle point approximation to get the leading order predictions. 

We treat fluctuations over the saddle point in \S \ref{fluctuations}, and show that saddle point analysis is valid above the upper critical dimension $d_u = 2r$ for higher order transitions. Afterwards, we formulate the Ginzburg criteria to classify when saddle point results are valid and when critical exponents are modified due to fluctuations below the upper critical dimension. The argument for the lower critical dimension remains unchanged as compared to the second order case. 

In \S \ref{secscaling} we obtain scaling forms of physical observables for higher order phase transitions from the renormalization group analysis of the system. We use this to derive scaling relations between critical exponents beyond the saddle point. We show that the imposition of analyticity in the phenomenological Hamiltonian combined with the scaling implies that critical exponents are the same above or below $T= T_C$ and $h=0$ (Appendix \ref{exponentspm}). We also show that the divergence of the correlation length naturally leads to the hyperscaling relation.

We then proceed onto computing corrections to the critical exponents using the one-loop beta functions in \S \ref{crrg}. We use this to calculate the critical exponents corresponding to $r=3$ and $r=4$ phase transitions below the upper critical dimension. We show that there are no relevant couplings in the vicinity of the non-trivial fixed point, and consequently, the sole corrections arise due to flows near the Gaussian fixed point. For $r \geq 5$, the one-loop corrections to the beta function vanish, and hence we need to go beyond one-loop calculations to find corrections to the critical exponents for such cases. 

 In general, there also exist phase transitions which are described by non-local order parameters. An important class of such phase transitions include transitions characterized by a gap in the eigenvalue spectrum. These include the Gross-Witten-Wadia (GWW) model \cite{Gross:1980he, WADIA1980403, wadia2012study}, Douglas-Kazakov model \cite{Douglas:1993iia}, Brownian walk models which map onto two-dimensional continuum Yang-Mills with different gauge groups \cite{FORRESTER2011500}, constrained Coulomb gas \cite{Cunden_2017}, bipartite entanglement \cite{PhysRevA.81.052324, PhysRevLett.101.050502, PhysRevLett.104.110501} and various other examples \cite{PhysRevLett.119.060601, PhysRevLett.107.177206, PhysRevE.88.042125, PhysRevLett.109.167203, 5730574, PhysRevLett.101.216809}. Further examples of third order phase transitions are given in the review \cite{Majumdar_2014}. Such phase transitions are outside the scope of our present work and will be dealt with in a forthcoming work.
 
 Fourth order phase transitions with local order parameter have been proposed as the nature of superconducting transition in certain materials in \cite{pardeep97, articlePK99,articlePK00, articlePK02, doi:10.1080/14786430802585158, doi:10.1080/13642810208223158, PhysRevB.71.104509}\footnote{We feel \cite{pardeep97, articlePK02, doi:10.1080/13642810208223158} have not received significant attention in the literature regarding higher order phase transitions, and in particular, we found their work only after we had independently rederived some of their results, and consequently there is overlap between parts of \S \ref{Landauth}, \S \ref{sad} and \S \ref{scalerg} of our work with theirs, which we have clearly outlined.}. Another example of fourth order phase transition is the Ising model on the Cayley tree proposed in \cite{STOSIC20091074}. Divergences using zeroes of the partition function in higher order phase transitions were analyzed in \cite{JANKE2006319}.

Apart from thermodynamic phase transitions, higher order phase transitions also occur in the context of topological phase transitions. An example of this is the phase transition between 2D Chern insulators, in which the third derivative of the free energy has a discontinuity. However, there are no local order parameters for such topological phase transitions, and consequently, Landau theory cannot be formulated to describe them. Consequently, we will not be looking at such phase transitions in our work.

\section{Higher order phase transitions with local order parameters: Landau theory}\label{sec1}
In this section, we will study a class of $r$th order phase transitions, which are described using local order parameters. We argue that the phenomenological theory describing these phase transitions in the infrared (IR) can be described using the Landau formalism. We will also introduce critical exponents corresponding to $r$th order phase transitions and calculate them using the saddle point approximation.

\subsection{The Landau Hamiltonian for higher order phase transitions}
\label{Landauth}

We work in $d$ spatial dimensions, which can be understood both as the non-relativistic spatial limit of a  $d+1$ dimensional relativistic theory or as an analytic continuation of the Lorentzian theory to a Euclidean theory via Wick rotation. Let us now write down the most general Hamiltonian, which obeys the following assumptions:  
\begin{enumerate}
     \item
      \textbf{Analyticity assumption}: We assume that the critical exponents are analytic everywhere apart from a singular line which terminates at the critical point in the plane spanned by the external field $h^i$ and temperature difference $t$ (where $t = \frac{T- T_C}{T_C}$), with the line given by $h^i=0$ and $t<0$ as shown in Figure \ref{fig:ht}. In other words, the free energy is analytic everywhere in the plane apart from this line. Thus the free energy term containing the external field in the phenomenological Hamiltonian should be chosen such that the associated critical exponents are analytic, i.e., the critical exponents have the same values when the critical point is approached from different directions in the $(h,t)$ plane. 
      
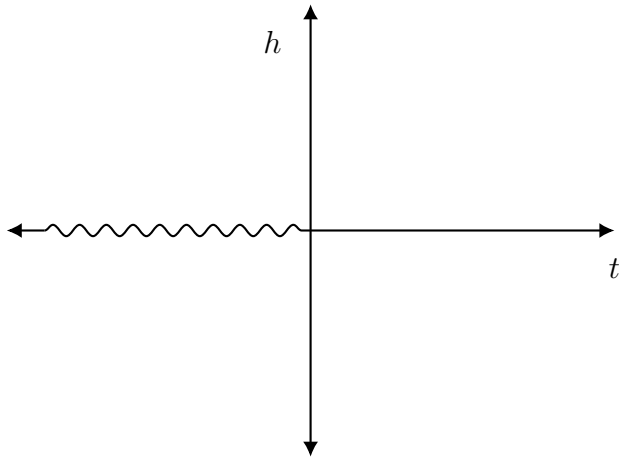
\begin{figure}[h]
\begin{center}
\begin{tikzpicture}
\draw[black, thick , <->] (0,-3) -- (0,3);
\draw[black, thick , ->] (0,0) -- (4,0);
\draw[black, thick, decorate, decoration={snake, amplitude= 0.75mm}] (-3.5,0) -- (0,0);
\draw[black, thick , ->] (-3.5,0) -- (-4,0);

 \node at (-0.5,2.5) {$h $};
  \node at (4,-0.5) {$t $};
\end{tikzpicture}
\end{center}
\caption{The order parameter $\phi$ is analytic everywhere except on the branch cut in $(h-t)$ plane shown above.} 
\label{fig:ht}
\end{figure}

      \item
      \textbf{Assumption regarding higher order terms}: We assume a small external field $h$ such that only the leading order term in $h$ contributes and higher order terms are suppressed. We also demand that terms involving higher derivatives of the order parameter are suppressed. These conditions are weaker than the first condition and serve to simplify our calculations and to clearly demonstrate the physics of the problem at hand.
      \item 
    \textbf{Fine tuning assumption}: Let $\phi_i(x)$ denote the order parameter of the phase transition. In hindsight, for $r$th order phase transitions, we assume that the terms $\abs{\phi_i}^{2(k-1)}, \, k < r$ either do not appear in the Hamiltonian, or their coefficients have a very small magnitude and do not change signs at the critical temperature of the phase transition. We enforce this requirement so that these terms do not alter the order of the phase transitions, as a significant contribution from such terms can potentially change the order. Similarly, we also demand that the terms of the form $\abs{\phi_i}^{2k}, \, k > r$ die off for a $r$th order phase transition. 
    \item
    \textbf{Rotational symmetry}: We further assume that our system possesses rotational symmetry and hence the Hamiltonian is invariant under O($N$) transformation ($N\geq 1$) given by
    \beq
    \phi_i \to \phi_i' = R_i^j \, \phi_j, \quad R \in O(N)
    \eeq
    where the rotation matrices $R$ belong to the matrix representation of $O(N)$.
\end{enumerate}

The most general phenomenological IR Hamiltonian obeying above assumptions which describes $r$th order phase transition is given by:
\begin{equation}
 \beta H = \int d^dx \left[\frac{K}{2} \abs{\grad \phi_i}^2 +   t_r \,|\phi_i (x)|^{2(r-1)} + u_r \, |\phi_i (x)|^{2r}   - (h_i\phi^i) \, |\phi_i (x)|^{2(r-2)}\right], \label{LG1}
\end{equation}
where we are using Einstein summation notation to sum over the index $i$ i.e. $|\phi_i|^2 = \sum_{i=1}^{n} \phi^i \phi_i$. Here $K$ is the coefficient of the kinetic term and the couplings $t_r$ and $u_r$ are expressed as functions of temperature as follows: 
\beq\label{2.3}
\begin{split}
t_r (T) &= c(T- T_C) + \text{O}(T- T_C)^2 = c't + \text{O}(t^2)  \\
u_r (T) &= u_0 + u_1(T- T_C) + \text{O}(T- T_C)^2 = u_0 + u_1' t + \text{O}(t^2)
\end{split}
\eeq
where $T_C$ is the temperature at which the transition takes place, $c, c', u_1, u_1'$ and $u_0$ are undetermined constants, and $t = \frac{T- T_C}{T_C}$ as defined previously. The constant $u_0$ must be positive in order to ensure that our Hamiltonian is positive-definite while $c$ must be positive in order to obtain the correct description of phases above and below the critical temperature. The exact values of these constants depend on the details of the system. The Hamiltonian in \eqref{LG1} was also discussed for the specific case of third and fourth order superconducting phase transitions in \cite{pardeep97, articlePK99,articlePK00, articlePK02, doi:10.1080/14786430802585158, doi:10.1080/13642810208223158, PhysRevB.71.104509}. 

The mass dimensions of the couplings appearing in \eqref{LG1} are given by:
\beq\label{couplingdimensions}
\begin{split}
[t_{r}] &= 2(r + d-1) - dr\\
[u_{r}] &= r(2 - d) +d\\
[h] &= d + (2r-3)\lc 1 - \frac{d}{2}\rc .
\end{split}
\eeq
Notice that the couplings in the above Hamiltonian can be non-renormalizable depending on the dimension $d$, and hence the physical observables can potentially be plagued by UV divergences untameable  by a finite number of counterterms. However our Hamiltonian is only defined in the IR, and we can impose a UV cutoff upto which the Hamiltonian description is valid. Thus we are not bothered by non-renormalizable interactions, analogous to the standard treatment of second order phase transitions in $d>4$.

 The (Gibbs) partition function of the system described by the Hamiltonian \eqref{LG1} is given by:
\begin{equation}
 Z = \int [D\phi(x)] \, \exp\left[ - \beta H(\phi(x), \, h)\right]. \label{LG2}
\end{equation}
The (Gibbs) free energy is given by $\beta F = - \log Z \approx V_d \, \text{ min} \left[ V(\phi) \right]_{\phi_0}$ where $V_d$ is the $d$-dimensional volume and the potential $V(\phi)$ is evaluated at the minima $\phi_0$. 

\subsection{Critical exponents}
\label{seccrit}
In this subsection, we define various critical exponents and compute them using saddle point approximation. We will drop the index $i$ in $\phi_i$ for notational convenience at various places in the text.
\subsubsection{Definition}
Let us introduce the critical exponents corresponding to these phase transitions. These exponents are generalizations of the exponents which characterize the second order phase transitions.

\begin{enumerate}
    \item The exponent\footnote{The critical exponent $\beta$ should not be confused with $\beta$ appearing in the partition function, which is given by $\beta = \frac{1}{k_B T}$. To avoid confusion, whenever we talk about the critical exponent $\beta$, we mention it explicitly.} $\beta$ is defined as the divergence of the order parameter at zero external field:
    \beq \label{critbeta}
    \phi(t, h =0) \propto \begin{cases} 0, &\quad T>T_C; \\
    \abs{t}^{\beta}, &\quad T < T_C.\\
    \end{cases}
    \eeq
    
    \item The exponent $\delta$ is defined as the divergence of the order parameter at the critical temperature: 
    \beq \label{eqnofstate}
    \phi(T = T_C, h) \propto h^{\frac{1}{\delta}}. 
    \eeq
    
    \item Denoting $\Phi$ as the integral of the order parameter over the full space i.e. $\Phi = \int d^dx \, \phi(x)$, we define the generalized susceptibility (response function) as
    \beq
    \chi = \frac{\partial \braket{\Phi}}{\partial h} \Big|_{h=0}, \label{susceptibility}
    \eeq
   where $\braket{\Phi}$ is the one-point function defined as:
   \beq
   \braket{\Phi} = \frac{1}{Z}\int [D\phi(x)] \, \int d^dx \, \phi(x)\, \exp\left[ - \beta H(\phi(x), \, h)\right]
   \eeq
   and where $Z$ is given in \eqref{LG2}. The divergence of the susceptibility is given by 
    \beq \label{critgamma}
    \chi(T, h=0) \propto \abs{t}^{-\gamma}.
    \eeq
    Using the definition of susceptibility in equation \eqref{susceptibility}, we obtain it's expression in terms of the order parameter by differentiating the partition function w.r.t $h$. 
    \beq \label{meqn}
    \begin{split}
    \chi &\equiv \beta  \braket{\Phi (x) \, \Phi^{2r-3}(y)}_c = \beta\lc\braket{\Phi (x) \, \Phi^{2r-3}(y)} - \braket{\Phi (x)} \, \braket{\Phi^{2r-3}(y)}\rc \\
    &\equiv \beta \int d^dx \, d^dy \braket{\phi (x) \, \phi^{2r-3}(y)}_c  =\beta \int d^dx\, d^dy \lc \braket{\phi (x) \, \phi^{2r-3}(y)} - \braket{\phi (x)} \, \braket{\phi^{2r-3}(y)} \rc
    \end{split}
    \eeq
    where $\braket{\Phi (x) \, \Phi^{2r-3}(y)}_c$ and $\braket{\phi (x) \, \phi^{2r-3}(y)}_c$ are the connected correlators. Note that setting a UV cuttoff removes divergences associated with the evaluation of product of fields coinciding at a point $y$ in $\braket{\phi^{2r-3}(y)}$, thus making it a well-defined quantity.

    \item  
    The (Gibbs) free energy is given by $\beta F = - \log Z$. Let us denote the $n$th derivative of the free energy as $C_n = \frac{\partial^{n}}{\partial t^n}  \lc \frac{\beta F}{V}\rc$. Since we are interested in a $r$th order phase transition, we expect the $r$th derivative of the free energy to diverge at $t = 0$. We define the exponent $\alpha$, which characterizes this divergence, as follows: 
    \beq \label{gap1}
    C_r \propto \abs{t}^{-\alpha}.
    \eeq
    Note here that the specific heat is given by the second derivative of free energy i.e. $ C_V = -C_2 \propto  \frac{\partial^{2}}{\partial t^2}  \lc \frac{\beta F}{V}\rc$. Hence for a second order phase transition, $C_V$ diverges. For $r$th order phase transition, the specific heat is given by:
    \beq \label{gap2}
    C_V(T, h=0) \propto \abs{t}^{r-\alpha -2}.
    \eeq
    
    \item  We also define a two-point connected correlator given by
    \beq \label{gfeqn}
    G^{r}_c(x,y) = \braket{\phi (x) \, \phi^{2r-3}(y)}_c = \braket{\phi (x) \, \phi^{2r-3}(y)} - \braket{\phi (x)} \, \braket{\phi^{2r-3}(y)}
    \eeq
     Note that $G^{r}_c(x,y)$ only depends on the separation between the operator insertions. This is because we are working in Euclidean space which has translations and rotations as a part of its isometry group.  If the separation is taken to be very large than the corelation length $\zeta$ i.e. $\abs{x-y} \gg \zeta$, then $G^{r}_c(x,y)$ roughly decays as $G^{r}_c(x,y) \sim \exp{-\frac{\abs{x-y}}{\zeta}}$.

     We can also write $\chi$ as given in equation \eqref{meqn} as an integral over the connected Green function as defined in equation \eqref{gfeqn}
    \beq \label{gap3}
    \chi = \beta \int d^d x \, d^d y \, G^{r}_c(x,y) = \beta V \int d^d\rho \, G^{r}_c(\rho,0), 
    \eeq
    where we have defined $\rho = \abs{x-y}$. Let the largest value of $G^{r}_c(\rho,0)$ be given by $G_0$ when $\rho < \zeta$, where $G_0$ is always finite.  Using the mean value theorem, we have the following inequality, where $V$ is the system's volume (where $V \geq \zeta^d$):
    \beq \label{divergg}
    \frac{\chi}{\beta V} < G_0\, \zeta^d.
    \eeq
    Here we have assumed that the contributions from $\rho \gg \zeta$ are negligible since the correlator dies off exponentially. Thus when $\chi$ diverges, the above inequality ensures that $\zeta$ necessarily diverges as well. This gives rise to the critical exponent $\nu$ defined by
    \beq
    \zeta (T, h=0) \propto \abs{t}^{-\nu}.
    \eeq

\end{enumerate}
    Note that here we have not labelled two different exponents $\gamma_{\pm}, \, \alpha_{\pm}$ etc. while approaching from the $t<0$ and $t>0$ directions. This is because we have already imposed analyticity in the $(h,t)$ plane while writing the phenomenological Hamiltonian, and consequently, we have $\gamma = \gamma_{\pm}, \, \alpha = \alpha_{\pm}$ and so on. It can be shown that even for deviations away from the saddle point, we still get single-valued exponents in the $(h,t)$ plane. We will prove this statement in Appendix \ref{exponentspm}.
    
    \subsubsection{Dilatation symmetry at critical temperature}
    \label{dilatation}
In the previous subsection, we discussed that at critical temperature, the correlation length $\zeta$ diverges as a consequence of the divergence of susceptibility, which follows from equation \eqref{divergg}. At the critical temperature, the correlation length becomes infinite (or more precisely, it becomes of the order of system size). Since there are no other length scales in the problem, $G^r_c(x,y)$ should decay as a power law. This implies that there is an emergent dilatation symmetry at the critical temperature, and the connected Green function transforms as follows
\beq
G_c^r \lc t =0, \lambda x \rc = \lambda^{\kappa} \, G_c^r \lc t =0, x \rc.
\eeq
Here $\kappa$ is the scaling dimension under the scaling transformation. Thus the system at criticality exhibits self-similarity at all intermediate scales between system size and the UV cutoff. 

\subsection{Critical exponents from saddle point approximation}\label{sad}
In this subsection, we will compute the critical exponents defined above, using saddle point approximation. The saddle points can be found by minimizing the potential appearing in the Hamiltonian \eqref{LG1}. The potential is given by (as mentioned before, we have dropped the index $i$ for notational convenience):
\begin{equation}
 V\ls \phi\equiv \phi(x)\rs =  t_r \, \phi^{2(r-1)} + u_r \, \phi^{2r} - (h.\phi) \, \phi^{2(r-2)}. \label{LG4}
\end{equation}
 
 \begin{enumerate}
    \item 
Working in the saddle point approximation, we impose $\frac{\partial V}{\partial \phi} = 0$ to get the values for the order parameter above and below the critical temperature,
\beq \label{msaddle}
\phi = \begin{cases}\sqrt{\dfrac{-t(r-1)}{ru}}, \quad &t<0\\
0, \quad &t>0.
\end{cases}
\eeq

This reduces to the familiar value $\phi^2 = \frac{-t}{2u}$ for second order phase transitions. For the special case of $r=3, 4$, equation \eqref{msaddle} matches with the expressions derived in \cite{pardeep97}. Comparing this with \eqref{critbeta}, we obtain $\beta = 1/2$.
   \item
Upon substituting the value of $\phi$ from eqn \eqref{msaddle}, the free energy $\frac{\beta F}{V} = \text{min}  \ls V(\phi)\rs $ is given by
\beq
\frac{\beta F}{V} = \begin{cases} - \lc \dfrac{u}{r-1}\rc\lc\dfrac{-t(r-1)}{ru}\rc^{r}, \quad &t<0\\
0, \quad &t>0.\\
\end{cases}
\eeq

For the special case of $r = 3,4$; this matches with the expression of free energy derived in \cite{pardeep97}. Thus the $r$th derivative of the free energy with respect to the temperature is discontinuous at $T=T_C$, which is given by
\beq \label{CR}
C_r(T, h=0) \propto \begin{cases}(-1)^{r+1}\lc\dfrac{r! \, u\, (r-1)^{r-1}}{(ur)^{r}}\rc , \quad &t<0\\
0, \quad &t>0\\
\end{cases}
\eeq

For $r=2$, this reduces to the familiar value for the specific heat, i.e. $C_2 = - C_V \propto \frac{1}{2u}$. Thus using \eqref{CR} we conclude that $\alpha = 0$.

\item
Let us now look at the critical exponent associated with susceptibility. Using our Hamiltonian and the definition of susceptibility from \eqref{susceptibility}, we obtain the following expression.
\beq
\frac{1}{\chi} = \begin{cases} -\dfrac{4(r-1)t}{(2r-3)}, \quad &t<0\\
-\dfrac{2(r-1)t}{(2r-3)}, \quad &t>0
\end{cases}
\eeq
This gives us $\gamma = 1$.

\item
The critical exponent $\delta$ can be obtained from setting $t=0$ in the saddle point equation of state:
\beq
h(\phi)= \frac{2\phi^3ru}{(2r-3)}.
\eeq
Using \eqref{eqnofstate}, we obtain $\delta = 3$.
\end{enumerate}

\begin{table}[h!]
\begin{center}
\begin{tabular}{|l|l|l|l|l|l|l|} 
\hline
Critical Exponents & $\alpha$ & $\beta$ & $\gamma$ & $\delta$ & $\nu$ (See \S\ref{secscaling}) & $\Delta$ (See \S \ref{secscaling} ) \\
  \hline
   &      &    & && &  \\
Landau Theory  & 0     &   $\dfrac{1}{2}$ &1 &3& $\dfrac{1}{y_t} = \dfrac{1}{2(r + d-1) - dr}$;  &$\dfrac{y_h}{y_t} = \dfrac{d + (2r-3)\lc 1 - \frac{d}{2}\rc}{2(r + d-1) - dr}$;                       \\
(Saddle point) &      &    & &&if \, $y_t$ is relevant & if \, $y_t$ is relevant \\
\hline
\end{tabular}
\caption{Critical exponents from the saddle point expansion in Landau theory.}
\label{table1}
\end{center}
\end{table}
We will introduce another exponent $\Delta$ and $\nu$ in \S \ref{secscaling}. The critical exponents from Landau theory are displayed in Table \ref{table1} for convenience.

\section{Fluctuations over saddle point approximation}
\label{fluctuations}
In this section, we will venture beyond the saddle point analysis by considering small fluctuations about the saddle point and taking their effects into account. Readers who are not interested in the calculational aspects of fluctuations can read the following introductory discussion of this section where we have stated the main results and skip directly to \S \ref{Ginzburg}.

We will show that incorporating fluctuations about the saddle point into the analysis of $r$th order phase transitions leads to the following features:

\begin{enumerate}
    \item Fluctuations which lead to powers of the order parameter in the Hamiltonian of the order ${2(k-1)}, \, k < r$  destroy the delicate analytic properties of the critical exponents. We will show this by looking at the case of quadratic fluctuations, where we find such fluctuations destroy analyticity for $r>2$.
    \item In $d< 2r$, saddle point analysis does not always hold. The Ginzburg criteria as defined in \S \ref{Ginzburg} tells us when saddle point analysis holds in $d< 2r$, and when fluctuations contribute significantly. In case fluctuations dominate, critical exponents are modified. 
    \item Saddle point analysis is valid above the upper critical dimension given by $d>d_u = 2r$. Hence the critical exponents corresponding to observables like $C_r$ do not change for $d> 2r$.
    \item Fluctuations destroy all order in the system for $d \leq d_l$ for second order phase transitions, where $d_l$ is the lower critical dimension  \cite{PhysRevLett.17.1133, Coleman:1973ci}. This argument remains unchanged for $r$th order phase transition as shown in Appendix \ref{kf}, where we show that $d_l = 2$ for systems with continuous symmetry (in the phenomenological Lagrangian) and $d_l =1$ for systems with discrete symmetry.
\end{enumerate}

 We begin by looking at the effect of fluctuations on two-point correlators. Afterwards, we analyze the effects of fluctuations on $C_r$. Specifically, we study how these observables' divergences near critical temperature are modified due to fluctuations. The computations in Appendix \ref{kf} also stress the importance of fluctuations, as we show that fluctuations lead to partial or complete destruction of order depending on the dimension of the system.

\subsection{Two point correlator and exponents}
 We are interested in fluctuations of the form:
\beq
\phi(x) = \lc \phi_0 + \phi_{\parallel}(x), \phi_{\bot,2}(x), \,  \dots \, , \phi_{\bot,n}(x)   \rc
\eeq
where $\phi_{\parallel}(x)$ and $\phi_{\bot,i}$ denote the longitudinal and transverse fluctuations respectively. The kinetic term goes as
\beq \label{derivative}
\lc \grad \phi \rc^2 = \lc \grad \phi_{\parallel} \rc^2 + \lc \grad \phi_{\bot} \rc^2,
\eeq
and the square of the order parameter goes as
\beq \label{sq}
\phi^2 = \phi_0^2 + 2 \phi_0 \phi_{\parallel} + \phi_{\parallel}^2 + \sum_{i = 2}^{n}\phi_{\bot,i}^2
\eeq
By using \eqref{derivative} and \eqref{sq} in \eqref{LG1} and keeping upto only quadratic terms in the fluctuations, we obtain: 
\beq
\begin{split}
\beta H =& \int d^d x\lc t_r \phi_0^{2(r-1)} + u_r \phi_0^{2r}\rc + \int d^d x\ls \frac{K}{2} \lc \grad \phi_{\parallel} \rc^2 + \lc \frac{2K}{2\xi_{\parallel}^2}  \rc \phi_{\parallel}^2\rs\\
& + \sum_{i = 2}^{n} \int d^d x\ls \frac{K}{2} \lc \grad \phi_{\bot,i} \rc^2 + \lc \frac{2K}{2\xi_{\bot,i}^2}  \rc \phi_{\bot,i}^2\rs + \text{O}\lc \phi_{\parallel}^3 , \phi_{\bot}^3 \rc \\
\end{split}
\eeq
where $\xi_{\parallel}$ and $\xi_{\bot,i}$ are correlation scales of the fluctuations and the first term is the potential evaluated at the saddle point. The correlation lengths $\xi_{\parallel}$ and $\xi_{\bot,i}$ corresponding to the two point correlator are explicitly given by:
\beq \label{xi0}
\begin{split}
\frac{\partial^2 V}{\partial \phi_{\parallel}^2} \bigg|_{\phi_0} &= \dfrac{2K}{\xi_{\parallel}^2}  = \begin{cases} 2ru_r \, \lc \dfrac{t-rt}{ru}\rc^{r-1} , \quad &t< 0 \\
t(r-1)(2r-3) \, \delta_{r,2}\quad &t> 0 \\\end{cases}  \\
\frac{\partial^2 V}{\partial \phi_{\bot,i}^2} \bigg|_{\phi_0} &= 
\frac{2K}{\xi_{\bot,i}^2} = \begin{cases} 0 , \quad &t< 0 \\
t(r-1) \, \delta_{r,2}\quad &t> 0 \\\end{cases}
\end{split}
\eeq
Note that the lengths $\xi_{\parallel}$ and $\xi_{\bot,i}$ corresponding to the two point correlator display a different analytic behaviour as compared to $\zeta$. The two point correlators $\xi$ which were zero in the saddle point approximation acquire a non-zero value due to fluctuations. It can be seen from the nature of the longitudinal polarization that the two point correlator $\xi$ is not an analytic function, since quadratic terms in the Hamiltonian characterizing the fluctuations destroy the delicate analytic features. As a result we have the scalings:
\beq \label{xi}
\begin{split} 
    \xi_{\parallel}^+ \sim \abs{t}^{-\frac{1}{2}} \delta_{r,2}, \quad \quad & \xi_{\parallel}^- \sim \abs{t}^{\frac{1-r}{2}} \\
   \xi_{\bot,i}^+ \sim \abs{t}^{-\frac{1}{2}} \delta_{r,2}, \quad \quad  & \xi_{\bot,i}^- \sim 0\\
\end{split}
\eeq

We note from equation \eqref{xi} that only in the case of second order phase transitions with $r=2$ we have $\xi_{\parallel}^+ = \xi_{\parallel}^- \equiv \xi_{\parallel} = \abs{t}^{-\frac{1}{2}}$. This is because such fluctuations give rise to analytic terms in the Hamiltonian corresponding to the second order phase transitions. 

Here we have shown that quadratic fluctuations destroy the analyticity of the physical observables in the $h-t$ plane. The same argument can be extended to fluctuations of the order ${2(k-1)}, \, k < r$.

\subsection{Fluctuation corrections to the saddle point}
Let us now evaluate the effect of fluctuations on the full partition function. By including the quadratic fluctuations in the partition function we obtain:
\beq
\begin{split}
Z \approx  \exp &\ls -V_d V(\phi_0)\rs  \int \ls D\phi_{\parallel}\rs \exp \ls -\frac{K}{2} \int d^dx \ls\lc \grad \phi_{\parallel}\rc^2 + \frac{2 \phi_{\parallel}^2}{\xi_{\parallel}^2} \rs\rs\\
\times & \int \ls \prod_{i} D\phi_{\bot,i}\rs \exp \ls -\frac{K}{2} \int d^dx \ls\lc \grad \phi_{\bot,i}\rc^2 + \frac{2 \phi_{\bot,i}^2}{\xi_{\bot,i}^2} \rs\rs
\end{split}
\eeq
where $V_d$ is the volume of the system. The free energy of the system is given as follows (where we have removed the label $i$ indicating the transverse directions for convenience):
\beq\label{freeen}
f = -\frac{\ln Z}{V_d} = V(\phi_0) + \frac{1}{2} \int \frac{d^dq}{(2\pi)^d} \ln \ls K \lc q^2 + 2 \lc \xi_{\parallel}\rc^{-2}\rc\rs + \, \frac{n-1}{2} \int \frac{d^dq}{(2\pi)^d} \ln \ls K \lc q^2 + 2 \lc \xi_{\bot}\rc^{-2}\rc\rs, 
\eeq
where $q$ is the momentum associated with the fluctuations. We will now look at the effect of fluctuations on the singularity structure of the $r$th derivative of the free energy which is given by $C_r \equiv \frac{d^r f}{dt^r}$. Using  \eqref{xi} and \eqref{freeen} we obtain the following corrections to $C_r$:
\beq
C_r \propto   \begin{cases} C_r^+ + \bigintss \dfrac{d^dq}{(2\pi)^d}  \dfrac{(-2)^{r-1}(r-1)! \delta_{r,2}}{\lc q^2 + 2 (\xi_{\parallel}^+)^{-2}\rc^r}   + (n-1) \bigintss \dfrac{d^dq}{(2\pi)^d}  \dfrac{(-2)^{r-1}(r-1)!\delta_{r,2}}{\lc q^2 + 2 (\xi_{\bot}^+)^{-2}\rc^r}  ; \quad  \quad & t>0 \\
 C_r^- + \bigintss \dfrac{d^dq}{(2\pi)^d} \lc \dfrac{(-2)^{r-1}(r-1)! (r-1)^r (\xi_{\parallel}^-)^{\frac{2r(2-r)}{r-1}}}{\lc q^2 + 2 (\xi_{\parallel}^-)^{-2}\rc^r} + O\lc\frac{1}{q^{2r-1}}\rc\rc; \quad  \quad &t<0
\end{cases}
\eeq
where we have only written the most divergent correction to $C_r^-$ explicitly and $O\lc\frac{1}{q^{2r-1}}\rc$ denotes the subleading corrections. Notice that the following momentum-integral appears in the correction terms:
\beq
I = \int \dfrac{d^dq}{(2\pi)^d} \lc \dfrac{1}{q^2 + 2 (\xi)^{-2}}\rc^r
\eeq
where we have suppressed the labels indicating longitudinal or transverse, and whether $t$ is greater or less than 0. The integral $I$ has mass dimension $d-2r$, and thus diverges for $d \geq 2r$. Introduction of a UV regulator $\Lambda = \frac{1}{a}$ cuts these divergences off. We can perform a similar analysis for the integrals at $t<0$. We can thus write $C_r$ in $d> 2r$ as

\beq \label{dgcr1}
C_r \propto \begin{cases} C_r^+ +  A_1 \, n \, a^{2r-d}\, \delta_{r,2} &t>0\\
 C_r^- + A_2 \, a^{2r-d} + O(a^{2r-d-1})\quad &t<0
\end{cases}
\eeq
while $C_r$ in $d< 2r$ takes the following form upon rescaling $q$ by $ \xi^{-1}$ 

\beq \label{dgcr2}
C_r \propto \begin{cases} C_r^+ + A_3 \lc \lc \xi^+_{\parallel} \rc^{2r-d} + (n-1)\lc \xi^+_{\bot} \rc^{2r-d} \rc \delta_{r,2} & t>0 \\
 C_r^- + A_4 \lc \xi^-_{\parallel} \rc^{2r-d + \frac{2r(2-r)}{r-1}}  \quad &t<0
\end{cases}
\eeq
Note that in the equations above, $A_1, A_2, A_3$ and $A_4$ are constants, which do not influence the discontinuity.\\

\medskip

\textbf{Physical aspects of fluctuations:} We will now analyze the physical aspects of these computations. As shown in equation \eqref{dgcr1}, for $d>2r$, $C_r$ gets an additive constant correction proportional to $a^{2r-d}$ which does not introduce any new singularity in the expression. However, we see from \eqref{dgcr2} that in $d<2r$, there are corrections that have associated singular parts, as the correlation lengths diverge near critical points. The calculation above is done by taking into account the first order corrections; inclusion of higher order terms can potentially introduce more singular terms to $C_r$. Thus the saddle point analysis does not work correctly in $d< 2r$ because fluctuation contributions tamper with the predicted singularity structure. Therefore we denote $d= 2r$ as the upper critical dimension above which the saddle point approximation holds, and fluctuations do not modify the critical exponents. The following question naturally arises: Under what conditions does saddle point analysis robustly hold below the upper critical dimension where fluctuations modify the singularities?  We discuss this in the next subsection \S \ref{Ginzburg}.

The predicted corrections to $C_V$ match with the values for second order phase transitions with $r=2$. By substituting the expressions for $\xi$'s from equations \eqref{xi0} and \eqref{xi} into \eqref{dgcr2}, we also see that corrections to the saddle point critical exponent $\alpha$ of $C_r$ do not have the similar scaling for $t>0$ and $t<0$. As stated earlier, this arises due to the fact that quadratic contribution to the action destroys the analyticity of physical observables in the $h-t$ plane. Note that fluctuations for third order phase transitions were also studied in \cite{pardeep97} for a different Hamiltonian as compared to \eqref{LG1}, for which they concluded the upper critical dimension is given by $d=2$.

\subsection{The Ginzburg criterion for $r$th order phase transitions}
\label{Ginzburg}

In this subsection, we quantify when the effects of fluctuations dominate over the saddle point expectations. We do this by generalizing the Ginzburg criteria to understand when saddle point calculations accurately predict critical exponents in the presence of fluctuations. 

 In order that the saddle point prediction is the dominant contribution, we demand that the saddle point discontinuity $\Delta C_r \equiv C_r^+ - C_r^-$ is larger as compared to the contributions from fluctuations. Only in such situations, we can trust the saddle point computation. Using \eqref{xi} in \eqref{dgcr2}, we see that the the condition for saddle point analysis to dominate over fluctuations is given by:
\beq \label{GC1}
\begin{split}
\Delta C_r  \gg \Delta C_r^{\text{fluc.}} &= A_3 \lc \lc \xi^+_\parallel \rc^{2r-d} + (n-1)\lc \xi^+_\bot \rc^{2r-d} \rc \delta_{r,2} - \lc A_4 \lc \xi^-_{\parallel} \rc^{2r-d + \frac{2r(2-r)}{r-1}} \rc \\ &= A' \, \delta_{r,2} \, t^{-\lc\frac{2r-d}{2}\rc} + A'' t^{\lc\frac{d}{2}-r\rc(r-1) + r(r-2)} 
\end{split}
\eeq
where $A'$ and $A''$ are linear combinations of proportionality constants $A_3$ and $A_4$. Notice that the term $A'$ drops out for higher order phase transitions ($r>2$), whereas for $r=2$ both the terms have the same order of magnitude. Thus the requirement for saddle point singularities to dominate over the fluctuations is given by
\beq
\abs{t} \gg t_G \sim \lc \frac{A''}{\Delta C_r}\rc^{-\frac{2}{\lc d-2r\rc(r-1) + 2r(r-2)}}.
\eeq
For the case of $r=2$, we recover the standard Ginzburg criteria from equation \eqref{GC1} which is given by
\beq
\abs{t} \gg t_G \sim \lc -\frac{A'+A''}{\Delta C_V}\rc^{\frac{2}{4 -d}}.
\eeq
where we have used the fact that $C_2 = - C_V$.  The interpretation of this temperature scale is that it is possible to recover the saddle point critical exponents if the system does not go within a distance $t_G$ near the critical temperature $T_C$.

\section{Scaling relations among critical exponents}
\label{secscaling}
In the previous section, we saw that critical exponents obtain corrections due to fluctuations. We now address the following question: Given that fluctuations modify the saddle-point calculated values of the critical exponents, can we still find some robust relations between the critical exponents that are valid beyond the saddle point? As we already know, there exist scaling relations between the exponents for second-order phase transitions. Can we generalize them for $r$th order phase transitions? 

Let us begin by looking at the free energy obtained using saddle point description. It is given by
\beq
f (t,h) \propto \begin{cases} \dfrac{t^r}{u^{r-1}}, \quad &h = 0; t< 0\\
\dfrac{h^{\frac{2r}{3}}}{u^{\frac{2r}{3}-1}} \quad &h \neq 0; t = 0.
\end{cases}
\eeq
 Thus we can now write the above free energy (at saddle point) as a homogenous function of $t$ and $h$. 
\beq \label{scalingfe1}
\begin{split}
f(t,h) &= \abs{t}^{r} g\lc \frac{h}{t^{\Delta}}\rc=\abs{t}^{r} g(x), \quad \text{where} \quad x \equiv \frac{h}{\abs{t}^{\Delta}}
\end{split} 
\eeq
Here $\Delta$ is the generalization of the familiar gap exponent which appears for the case of second order phase transitions. The generalization of gap exponent for higher order phase transition was also discussed in \cite{JANKE2006319}. Let us now look at the $h \to 0$ and the $t \to 0$ limits of the function $g(x)$ such that the power of $h$ is kept intact. These go as
\beq \label{scalingfe2}
\lim_{x \to 0} g(x) \sim \frac{1}{u^{r-1}}, \quad \text{and} \quad \lim_{x \to \infty} g(x) \sim \frac{x^{\frac{2r}{3}}}{u^{\frac{2r}{3}-1}}.
\eeq
Equation \eqref{scalingfe2} implies that the free energy takes the following form near $h \neq 0$ and $t=0$:
\beq
f \sim \abs{t}^r \frac{h^{\frac{2r}{3}}}{t^{\frac{2r \Delta}{3}}\lc u^{\frac{2r \Delta}{3}-1}\rc}.
\eeq
Since $f$ remains finite at $t = 0$, the gap exponent is given by setting the coefficient of $\abs{t}$ to 0, and we obtain $\Delta = \frac{3}{2}$.

\subsection{Scaling form of observables from RG}\label{scalerg}
In this subsection, we will argue that beyond the saddle point, the singular part of the free energy has the following homogenous form:
\beq \label{fescaling}
f_{\text{sing}} (t, h) = \abs{t}^{r-\alpha} g\lc \frac{h}{t^{\Delta}}\rc 
\eeq
This form can be understood using Wilsonian Renormalization Group (RG) flow. Under the RG transformations, the partition function remains unchanged and hence the corresponding free energies are related by:
\beq
Vf(t ,h) = V' f(t_{b},h_b) 
\eeq
where $t_b$ and $h_b$ are the rescaled couplings obtained by rescaling $x \rightarrow x/b$, where $b > 1$. Notice that we have suppressed the subscript $r$ in $t_r$, which we will keep doing for the rest of this section as well. Hence, in $d$- dimensions, the rescaled volume gets smaller by a factor of $b^d$ and we obtain:
\beq
f(t,h) = b^{-d}f (b^{y_{t}} t, b^{y_h} h)
\eeq
where $y_t$ and $y_h$ are the mass dimensions of the couplings $t_r$ and $h$ as given in \eqref{couplingdimensions}. In case the coupling $t$ is relevant i.e. $y_t > 0$, we  can choose $b = t^{-1/y_{t}}$, such that $b>1$. Notice that if the Lagrangian does not contain any relevant coupling, this choice of $b$ is not possible and hence we cannot write down the scaling relations derived below. 
With above choice of $b$, the free energy takes the following form:
\beq\label{rgsc}
f(t, h) = t^{d/y_{t}}f \lc 1, \frac{h}{t^{y_h/y_{t}}}\rc \equiv t^{d/y_{t}} g_f \lc\frac{h}{t^{y_h/y_{t}}}\rc
\eeq
Thus we obtain the same form as \eqref{fescaling} with $r - \alpha = d/y_t$ and $\Delta =  y_h/y_t$. Notice that the above scaling is valid only when the coupling $t$ is relevant. 

Within the context of third and fourth order phase transition, a similar scaling relation for the free energy was discussed in \cite{doi:10.1080/13642810208223158, JANKE2006319}. We derive the above scaling using renormalization group and as described previously, the expression is valid only if the coupling $t_r$ in \eqref{LG1} is a relevant coupling within Wilsonian RG.

We can obtain similar homogeneous scaling forms for the specific heat and $C_r$ by differentiating $f$ w.r.t. $t$. We have:
\beq
\begin{split}
 \frac{df}{dt} &= (r - \alpha) \abs{t}^{r-\alpha -1} g\lc \frac{h}{t^{\Delta}}\rc  - \Delta h \abs{t}^{r- \alpha - \Delta- 1} g\lc \frac{h}{t^{\Delta}}\rc\\
 &= \abs{t}^{r-\alpha -1}\ls (r - \alpha) g\lc \frac{h}{t^{\Delta}}\rc  - \Delta\frac{h}{t^{\Delta}}  g\lc \frac{h}{t^{\Delta}}\rc\rs\\
 &= \abs{t}^{r-\alpha -1} \tilde{g} \lc \frac{h}{\abs{t}^{\Delta}} \rc
\end{split} 
\eeq
where $\tilde{g} \lc \frac{h}{\abs{t}^{\Delta}} \rc =  (r - \alpha) g\lc \frac{h}{t^{\Delta}}\rc  - \Delta\frac{h}{t^{\Delta}}  g\lc \frac{h}{t^{\Delta}}\rc$. Hence the specific heat takes the form
\beq \label{kekek}
C^{\text{sing}}_V \sim -\frac{\partial^2 f}{\partial t^2} \sim \abs{t}^{r-\alpha -2} \widehat{g}\lc \frac{h}{t^{\Delta}}\rc,
\eeq
while the $r$th derivative of the free energy takes the form
\beq
C^{\text{sing}}_r \sim -\frac{\partial^r f}{\partial t^r} \sim \abs{t}^{-\alpha} \bar{g}\lc \frac{h}{t^{\Delta}}\rc.
\eeq
\subsection{Identities satisfied by critical exponents}\label{relac}
In this subsection, we write down similar homogeneous forms for other physical observables near the critical point. We will use these scaling forms to obtain relations between various critical exponents. 
\begin{enumerate}
    \item Using the scaling of the free energy from eqn \eqref{fescaling} and the relation between the order parameter and the free energy, we conclude that $\phi$ scales as
\beq \label{mscaling}
\phi^{2r - 3} \sim |t|^{r-\alpha - \Delta} g_\phi \lc \frac{h}{t^{\Delta}}\rc, 
\eeq
using which we can conclude from equation \eqref{critbeta} that $\beta$ is given by:
\beq \label{betarelation}
\beta = \frac{r-\alpha -\Delta}{2r-3}.
\eeq
 \item
 When $x \to \infty$, the dominant contribution in $g_\phi$ scales as $g_\phi \sim x^p$ where $p$ is the leading power in the expansion, and therefore $\phi^{2r-3} \sim \abs{t}^{r-\alpha -\Delta} \lc \frac{h}{t^{\Delta}}\rc^p$. As this limit is $t$-independent, we have the following relation between the exponents
 \beq
 \Delta p = r-\alpha-\Delta, \quad \implies \quad \phi^{2r-3}(t=0, h) \sim h^{\frac{r-\alpha-\Delta}{\Delta}}.
 \eeq
 Using equation \eqref{eqnofstate} the above identity implies the following relation
 \beq \label{deltarelation}
 \delta = \frac{(2r-3)\Delta }{r - \alpha -\Delta} = \frac{\Delta}{\beta}.
 \eeq
 
 \item
 Using the definition of generalized susceptibility as $\chi = \frac{d \braket{\Phi}}{dh}$, and using the scaling of $\phi$ from eqn \eqref{mscaling}, we see that $\chi$ scales as
 \beq
 \chi(t,h) = \frac{d \braket{\Phi}}{dh} \sim \abs{t}^{\frac{r-\alpha -\Delta}{2r-3} -\Delta}  g_{\chi} \lc \frac{h}{t^{\Delta}}\rc.
 \eeq
 Using equation \eqref{critgamma} we obtain: \beq \label{gammarelation}
 \gamma = \Delta - \lc \frac{r-\alpha - \Delta}{2r-3}\rc.
 \eeq
\end{enumerate}

Thus we see that all the critical exponents $\lc \alpha, \beta, \gamma, \delta, \Delta\rc$ can be derived from two independent critical exponents (say from $\Delta$ and $\alpha$ in the case of second order phase transition). Using equations \eqref{betarelation}, \eqref{deltarelation} and \eqref{gammarelation}, we obtain the following higher-order generalizations of the Rushbrooke identity \cite{doi:10.1063/1.1734338} (the saturation of the Rushbrooke inequality), and the Widom scaling law \cite{doi:10.1063/1.1696618} for higher order phase transitions:
\beq
\alpha + 2(r-1)\beta + \gamma = r \qquad \& \qquad \delta -1 = \frac{\gamma}{\beta}
\eeq
Notice that the above identities are different from the identities derived in \cite{doi:10.1080/13642810208223158, JANKE2006319}. This is because \cite{doi:10.1080/13642810208223158} obtained their scaling relations by considering a different magnetic field coupling as compared to our coupling in the Hamiltonian given in \eqref{LG1}. As a consistency check for our above relations, we can set $r=2$ to obtain
\beq
\alpha + 2\beta + \gamma = 2 \qquad \& \qquad \delta -1 = \frac{\gamma}{\beta}
\eeq
which are the Rushbrooke's identity \cite{doi:10.1063/1.1734338} and Widom's scaling law \cite{doi:10.1063/1.1696618} respectively for the second order phase transitions. 

\subsection{Scaling relation from the divergence of correlation length}
In this subsection, we will use the two-point correlator given in equation \eqref{gfeqn} to understand it's divergence near the critical point. We hereby derive the generalized Josephson identity for $r$th order phase transitions. Following the main argument from our previous section, the correlation length $\zeta(t,h)$ has the following form:
\beq \label{23}
\zeta (t,h) \sim \abs{t}^{-\nu} g_{\zeta} \lc \frac{h}{\abs{t}^{\Delta}}\rc
\eeq
In the vicinity of the critical point, the most important length scale is $\zeta$, and the singular contribution to observables arise solely due to $\zeta$'s singularity. A similar singular behaviour was demonstrated in the case of quadratic fluctuations in \S \ref{fluctuations}, where the correlation length $\xi$ dictated the singular contributions in $d<2r$, with the only significant difference being that the corrections led to non-analytic behaviour. Hence the singular part of the free energy in terms of the correlation length $\zeta$ is given by:
\beq \label{eqn24}
V_d f(t,h) \sim \ln Z = B_1 \lc \frac{L}{\zeta}\rc^d + B_2 \lc \frac{L}{a}\rc^d ,
\eeq
where $B_1$ and $B_2$ denote non-singular quantities, $L$ denotes the system size, and $\Lambda = \frac{1}{a}$ is the UV cutoff. Since the singular part of the free energy arises from the first term in the above equation, we have
\beq \label{Json}
f (t,h) \sim \frac{\ln Z}{L^d} \sim \zeta^{-d} \sim \abs{t}^{d\nu} g \lc \frac{h}{t^{\Delta}}\rc
\eeq
This is the same scaling form as given in \eqref{rgsc} with 
\beq \label{nurel}
\nu = \frac{1}{y_{t}}\, , \qquad \qquad \Delta =\frac{y_h}{y_{t}}.
\eeq
The scaling form in \eqref{Json} has the following important properties:
\begin{enumerate}
    \item From \eqref{eqn24}, we observe that the free energy grows as $\lc \frac{L}{\zeta}\rc^d$ near the critical point. It can be interpreted as a measure of the system size divided by the correlation length $\zeta$, such that the system is divided into many small blocks. Each small block can be thought of as an independent degree of freedom which contributes a constant factor to the overall free energy. We also naturally obtain the homogeneity of the singular part of the free energy.
    
    \item
    We arrive at the generalized Josephson's identity by comparing \eqref{Json} with the homogenous form of free energy given in \eqref{fescaling}. The identity is given by
    \beq
    r - \alpha = d \nu.
    \eeq
    Josephson's identity is basically a hyperscaling relation that holds below and at the upper critical dimension, but it does not hold above the upper critical dimension. This is because the scaling form for $\zeta$ as given in equations \eqref{23} and \eqref{eqn24} does not hold above the upper critical dimension. Above the upper critical dimension $g_{\zeta}(x)$ is singular in $x$ \cite{363515} and can be shown to grow as $g_{\zeta} (x)\sim \frac{1}{x}$ using mean-field theory, which is singular when $x \to 0$.

\end{enumerate}

We have summarized the four independent relations between the critical exponents $\alpha, \beta, \gamma, \delta, \Delta$ and $\nu$ in Table \ref{table2}. Thus all the critical exponents can be written in terms of two independent ones, say in terms of $\gamma$ and $\Delta$. 
 \begin{table}[h!]
\begin{center}
\begin{tabular}{|l|l|l|} 
\hline
Serial No. & Exponents & Relation \\

  \hline

1. & $\beta, \alpha, \Delta$  & $\beta = \dfrac{r-\alpha -\Delta}{2r-3} $ \vspace{2pt}\\

\hline

2. & $\delta, \alpha, \Delta$  & $\delta = \dfrac{(2r-3)\Delta }{r - \alpha -\Delta}$ \vspace{2pt}\\

\hline

3. & $\gamma, \alpha, \Delta$  & $ \gamma = \Delta - \lc \dfrac{r-\alpha - \Delta}{2r-3}\rc$ \vspace{2pt}\\
\hline

4. & $ \alpha, \nu$  & $ r-\alpha = d\nu$ \vspace{2pt}\\
\hline
\end{tabular}
\caption{Independent relations between different critical exponents}
\label{table2}
\end{center}
\end{table}

\section{Critical exponents about fixed points of RG flow}\label{crrg}
 In the previous sections, we calculated the critical exponents using the saddle point approximation and derived relations between critical exponents away from the saddle point. In this section, we will use perturbative RG flow to find fixed points of the theory and use them to calculate the critical exponents. Specifically, we will be looking at the case when fluctuations dominate over saddle point contributions as quantified by the generalized Ginzburg criteria in $d \leq 2r$. In that case, the critical exponents are captured by relevant couplings about fixed points. We will perform linear stability analysis near these fixed points to calculate the critical exponents in their vicinity.

Let us denote the full coupling space of our theory by $S (t_r, u_r, \dots)$. A particular Hamiltonian ($H$) is described by couplings at a point in this space. RG flow involves rescaling the couplings and renormalizing the order parameter, thereby taking us from one point in this parameter space to another, i.e., $S_b \rightarrow R_b S$. Here we have denoted the action of RG flow by $R_b$, with $b$ being the rescaling parameter defined in \S \ref{scalerg}. The fixed points of these flows are defined as $R_b S^* = S^*$. Since we rescale the couplings in the RG procedure, the correlation length $\zeta^*$ should either go to zero or infinity at the fixed point. When $\zeta^* = 0$, the system is in a completely disordered phase, and when $\zeta^* \rightarrow \infty$, the system is in the ordered phase. 

We can now study the stability of the fixed points by linearizing the beta function near them. The subspace of irrelevant couplings is known as the basin of attraction since the flow in this subspace converges towards the fixed point. Since the correlation length diverges at the fixed point, it diverges at each point in the basin of attraction. This basin of attraction is basically the critical surface at which the phase transition takes place. The behavior of critical exponents near the phase transition is hence determined only by the set of relevant couplings. We will use them to compute the value of critical exponents beyond the saddle point approximation.                                               

\subsection{One-loop beta function}
In this subsection, we compute the one-loop beta functions for the couplings appearing in the Hamiltonian \eqref{LG1}. We follow the procedure described in \cite{skinner, Hollowood:2009eh}. The Euclidean action for $r$th order phase transition at zero external magnetic field is given by:
\beq
S_{\Lambda}[\phi_i, h = 0] = \int d^dx \, \left[\frac{1}{2}(\partial_\mu \phi^i \partial^\mu \phi_i) + V(\phi_i)\right]
\eeq
where $\Lambda$ is the UV cutoff and the potential is given by:
\beq
V(\phi_i) = t_r(\phi^2)^{r-1} + u_r (\phi^2)^r
\eeq
with $\phi^2 = \sum_{i= 1}^{N} \phi_i \phi^i$. We work with dimensionless couplings which are defined as follows:
\beq
t_r = \frac{g_{2(r-1)}}{(2r-2)!}\Lambda^{d-(r-1)(d-2)} \, , \qquad \qquad u_r= \frac{g_{2r}}{2r!}\Lambda^{d-r(d-2)}\, .
\eeq
In terms of these dimensionless couplings, the potential is given by:
\beq\label{pot1}
V(\phi_i) = \frac{g_{2(r-1)}}{(2r-2)!}\Lambda^{d-(r-1)(d-2)}(\phi^2)^{r-1} + \frac{g_{2r}}{2r!}\Lambda^{d-r(d-2)} (\phi^2)^r
\eeq 
 We will now use this potential to study RG flows using the Wilsonian formalism. This involves splitting the scalar field in terms of high energy and low energy modes and the subsequent procedure of systematically integrating out the high energy modes. Therefore we split the field $\phi^i$ into the low energy modes $\varphi^i$ and the high energy modes $\chi^i$ i.e. $\phi^i = \varphi^i + \chi^i$. Hence the action can be expanded as:
\beq
S[\phi^i] = S[\varphi^i] + \int d^d x \, \left[\frac{1}{2}(\partial_\mu \chi^i \partial^\mu\chi_i) + \frac{1}{2}\chi^i\frac{\partial^2V (\varphi^k)}{\partial\varphi^i\partial\varphi^j}\chi^j + .. \right]
\eeq
where we have chosen $\varphi^i$ such that it minimises the potential i.e. $V'(\varphi^i) = 0$. We will now integrate out the high energy modes $\chi^i$ in the partition function, by lowering the scale infinitesimally i.e. setting $\Lambda'$ = $\Lambda - \delta \Lambda$. We will focus on the first order terms in $\delta \Lambda$, which is similar to computing one-loop effects only. It can be argued that at this order, only the quadratic terms in $\chi^i$ contribute to the effective action \cite{skinner}. Upon integrating out the high energy modes, the one-loop correction to the effective action is given by:

\beq\label{effac}
\delta_\Lambda S[\phi] = \frac{\Lambda^{d-1} \delta\Lambda}{(4 \pi)^{d/2}\Gamma(d/2)} \int d^d x \, \textrm{ln} \bigg[\textrm{det}\left(\Lambda^2 \delta_{ij} + \frac{\partial^2V (\varphi^i)}{\partial\varphi^i\partial\varphi^j}\right)\bigg]
\eeq
This process of integrating out the high energy modes changes the effective couplings depending on the UV scale at hand. The change in the $k$th coupling is captured by the beta function $\lc \beta_k =  dg_k/d \log\Lambda\rc$. We can use \eqref{effac} to compute the beta functions of the couplings $g_{2(r-1)}$ and $g_{2r}$ appearing in the potential \eqref{pot1}, which are given by:
\begin{eqnarray}\label{beta}
\Lambda \frac{d g_{2(r-1)}}{d \Lambda} &=& [(r-1)(d-2)-d] g_{2(r-1)} - a \Lambda^{(r-1)(d-2)}\frac{\partial^{2(r-1)}}{\partial \varphi_i^{2(r-1)}}\bigg[ \textrm{ln} \bigg[\textrm{det}\left(\Lambda^2 \delta_{ij} + \frac{\partial^2V (\varphi^i)}{\partial\varphi^i\partial\varphi^j}\right)\bigg]\bigg]\bigg|_{\varphi^i = 0}\nonumber\\ 
\Lambda \frac{d g_{2r}}{d \Lambda} &=& [r(d-2)-d] g_{2r} - a\Lambda^{r(d-2)} \frac{\partial^{2r}}{\partial \varphi_i^{2r}}\bigg[ \textrm{ln} \bigg[\textrm{det}\left(\Lambda^2 \delta_{ij} + \frac{\partial^2V (\varphi^i)}{\partial\varphi^i\partial\varphi^j}\right)\bigg]\bigg]\bigg|_{\varphi^i = 0}
\end{eqnarray}
where we have defined $a$ as follows
\beq\label{a}
a = \frac{1}{(4 \pi)^{d/2}\Gamma(d/2)}.
\eeq
Using \eqref{pot1} we compute the first derivative of potential which is given by:
\begin{eqnarray*}
\frac{\partial V}{\partial \varphi^i} &=& \frac{g_{2(r-1)}}{(2r-3)!}\Lambda^{d-(r-1)(d-2)}\left(\sum_m\varphi_m^2\right)^{r-2}\varphi_k \delta^{ik} + \frac{g_{2r}}{(2r-1)!}\Lambda^{d-r(d-2)} \left(\sum_m\varphi_m^2\right)^{r-1}\varphi_k \delta^{ik},
\end{eqnarray*}
while the second derivative is given by:
\begin{eqnarray*}
\frac{\partial^2 V}{\partial \varphi^i \varphi^j} &=& \frac{g_{2(r-1)}}{(2r-3)!}\Lambda^{d-(r-1)(d-2)}\left(\sum_m\varphi_m^2\right)^{r-2}\delta^{ij}  +\frac{g_{2r}}{(2r-1)!}\Lambda^{d-r(d-2)} \left(\sum_m\varphi_m^2\right)^{r-1}\delta^{ij}\\
&& + 2(r-1)\frac{g_{2r}}{(2r-1)!}\Lambda^{d-r(d-2)} \left(\sum_m\varphi_m^2\right)^{r-2}\varphi_k\varphi_l\delta^{ik}\delta^{jl}\\
&& + 2(r-2) \frac{g_{2(r-1)}}{(2r-3)!}\Lambda^{d-(r-1)(d-2)}\left(\sum_m\varphi_m^2\right)^{r-3}\varphi_k\varphi_l\delta^{ik}\delta^{jl}. 
\end{eqnarray*}
We can now compute the determinant appearing in the expression for beta function \eqref{beta} which takes the following form:
\begin{eqnarray}\label{determinant}
\textrm{det}\left(\Lambda^2 \delta_{ij} + \frac{\partial^2V (\varphi^i)}{\partial\varphi^i\partial\varphi^j}\right) &=& \left[\Lambda^2 + \left(\sum_m\varphi_m^2\right)^{r-2}\left(\frac{g_{2(r-1)}\Lambda^{d-(r-1)(d-2)}}{(2r-3)!} + \frac{g_{2r}\Lambda^{d-r(d-2)}}{(2r-1)!}\sum_m\varphi_m^2\right)\right]^{N-1}  \nonumber\\
\times && \left[\Lambda^2 + \left(\sum_m\varphi_m^2\right)^{r-2}\left(\frac{g_{2(r-1)}\Lambda^{d-(r-1)(d-2)}}{(2r-4)!} + \frac{g_{2r}\Lambda^{d-r(d-2)}}{(2r-2)!}\sum_m\varphi_m^2\right)\right]\nonumber\\
\end{eqnarray}
Substituting the expression for the determinant given by \eqref{determinant} in the beta function \eqref{beta}, we can calculate the one-loop beta functions for various $r$. In the rest of this section, we compute the critical exponents for phase transitions of different orders.

\subsection{Critical exponents for second order phase transitions $(r=2)$}
\label{r2d2}
In this subsection, we focus on second order phase transitions ($r =2$). We can see that for $r =2$ the couplings $g_2$ and $g_4$ appear in the Hamiltonian \eqref{LG1}. We use \eqref{beta} to compute the one-loop beta-functions for these couplings and then evaluate the critical exponents for this case. The results in this subsection are a review of previously known results for second order phase transitions. Using \eqref{beta} and \eqref{determinant}, we find the following one-loop beta functions for the couplings $g_2$ and $g_4$:
\begin{eqnarray}
 \Lambda \frac{d g_{2}}{d \Lambda} &=& -2 g_{2} - \frac{(N + 2)}{(4 \pi)^{d/2}\Gamma(d/2)}\frac{g_{4}}{3(1 + g_2)}\\
 \Lambda \frac{d g_{4}}{d \Lambda} &=& (d- 4) g_{4} + \frac{(N+ 8)}{3(4 \pi)^{d/2}\Gamma(d/2)}\frac{g_{4}^2}{(1 + g_2)^2}
 \end{eqnarray}
 The fixed points of the RG flow can be found by setting $\beta(g_2)$ and $\beta(g_4)$ to zero. There are two fixed points. The first one is a trivial fixed point known as the Gaussian fixed point and it is given by:
 \begin{equation}
g_2^* = 0\, ,  \qquad \qquad  g_4^* = 0
\end{equation}
Near the Gaussian fixed point, the coupling $g_2$ is relevant and $g_4$ is marginal and their mass dimensions are given in \eqref{couplingdimensions}. There also exists a non-trivial Wilson-Fisher (WF) fixed point \cite{wf} given by:
\begin{equation}
g_2^* = -\frac{(d-4) (N+2)}{d (N+2)-6 (N+4)}\, ,  \qquad \qquad  g_4^* = -\frac{12 (d-4) (N+8)}{a (d (N+2)-6 (N+4))^2}
\end{equation}
where $a$ is defined in \eqref{a}. We will now study flows about this fixed point, which appears "just below" four dimensions. Defining $\epsilon \equiv 4-d$, the Wilson-Fisher fixed point is given by:
\begin{equation}
g_2^* = -\frac{\epsilon (N+2)}{2 (N+8)} + O(\epsilon^2)\, ,  \qquad \qquad  g_4^* = \frac{3 \epsilon}{a (N+8)}  + O(\epsilon^2)
\end{equation}
We can now perform $\epsilon$-expansion of the beta functions near these fixed points:
\begin{eqnarray*}
\beta(g_2^* + \delta g_2) = \beta(\delta g_2) &=& -2 (g_2^* + \delta g_2) - \frac{(N + 2)}{3a}\frac{(g_4^* + \delta g_4)}{(1 + g_2^* + \delta g_2)}\\
&=&  -2 \delta g_2 + \frac{\epsilon (N+2)}{(N+8)}\delta g_2 - \frac{a (N+2)}{3} \delta g_4 \left(1 +\frac{\epsilon (N+2)}{2(N+8)} \right) + O(\epsilon^2) \\
\beta(g_4^* + \delta g_4) = \beta(\delta g_4) &=& -\epsilon(g_4^* + \delta g_4) + \frac{a(N+ 8)}{3}\frac{(g_4^* + \delta g_4)^2}{(1 + g_2^* + \delta g_2)^2}\\
&=& \epsilon \delta g_4 + O(\epsilon^2)
\end{eqnarray*}
Hence we obtain the following matrix of the linearized beta functions about the Wilson-Fisher fixed point:
\beq
\begin{pmatrix}
   \beta(\delta g_2)\\
   \beta(\delta g_4)
\end{pmatrix} = \begin{pmatrix}
 -2 + \frac{\epsilon (N+2)}{(N+8)} \qquad &     - \frac{a (N+2)}{3} \left(1 +\frac{\epsilon (N+2)}{2(N+8)} \right) \\
 0 \qquad & \epsilon
\end{pmatrix}\begin{pmatrix}
   \delta g_2\\
   \delta g_4
\end{pmatrix}
\eeq
The eigenvalues\footnote{Note that our beta functions are negative as compared to \cite{kardar_2007, kardarcourse}, who perform renormalization by studying the variation of couplings with respect to length scale, as opposed to momentum scale in our case.} are $ -2 + \frac{\epsilon (N+2)}{(N+8)}$ and $\epsilon$. The first eigenvalue is negative and hence the coupling $g_2$ is relevant around the WF fixed point and the scaling dimension $y_t$ is given by:
\beq
y_t = 2 - \frac{\epsilon (N+2)}{(N+8)}
\eeq
On the other hand, the eigenvalue corresponding to the coupling $g_4$ is positive for $d<4$. Hence the coupling $g_4$ is irrelevant (making the WF fixed point stable along the corresponding eigenvector) for $d <4$. We can use $y_t$ to calculate the critical exponent $\nu$ using the relation \eqref{nurel} and it is given by:
\beq
\nu = \frac{1}{y_t} =  \lc 2\lc 1 - \frac{\epsilon (N+2)}{2(N+8)}\rc\rc ^{-1} = \frac{1}{2} + \frac{\epsilon}{4} \frac{ (N+2)}{(N+8)} + O(\epsilon^2)
\eeq
For $r =2$, the scaling dimension of the coupling $h_i$ is still given by \eqref{couplingdimensions}. This is because in the Lagrangian $h_i$ couples to the zero momentum mode $\int d^dx \, \phi_i(x) = \tilde{\phi}_i(k= 0)$, and hence integrating out high energy modes using Wilsonian RG leaves $h_i$ unchanged. Therefore for the special case of second order phase transition ($r = 2$) the scaling dimension of $h_i$ just below four dimensions is given by:
\beq
y_h = 3 - \frac{\epsilon}{2} + O(\epsilon^2)
\eeq
We can now calculate the gap exponent $\Delta$ from \eqref{nurel} which is given by:
\beq
\Delta = \frac{y_h}{y_t} = \frac{3}{2} + \frac{ \epsilon}{4} \lc\frac{3(N+2)}{N+8} - 1 \rc + O(\epsilon^2)
\eeq
We can use the exponents $\nu$ and $\Delta$ to compute all other critical exponents using relations derived in \S\ref{relac}. These exponents are given by:
\beq
\begin{split}
    \alpha &= \epsilon \lc\frac{1}{2} - \frac{N+2}{N+8}\rc + O(\epsilon^2)\\
    \beta &= \frac{1}{2} - \frac{\epsilon}{4} \lc1 - \frac{N+2}{N+8}\rc  + O(\epsilon^2)\\
    \gamma &= 1 + \frac{\epsilon}{2}\lc\frac{N+2}{N+8}\rc  + O(\epsilon^2)\\
    \delta &= 3 + \epsilon\lc \frac{N+2}{N+8}\rc + O(\epsilon^2)
\end{split}
\eeq
The above results match with the standard results for second order phase transitions as given in \cite{kardar_2007, kardarcourse, Peskin:1995ev}.
\subsection{Critical exponents for third order phase transitions $(r=3)$}
In this subsection, we calculate the critical exponents for $r=3$. The calculation of exponents for $r>2$ is exactly the same as for $r =2$ with an important distinction. For $r=2$, integrating out high energy modes does not renormalize the coupling $h_i$ because the term $h_i \phi_i$ is a zero momentum mode. Hence the mass dimension of the coupling $h_i$ does not receive any quantum corrections under Wilsonian renormalization. However we can see from going to the Fourier space that $h_i \phi_i^{2r-3}$ no longer has only zero-momentum contribution. This term can potentially become relevant in the IR and give large contributions.

In order to simplify the detailed calculation and qualitatively understand the underlying physics of the problem, we work in the limit when $h_i \phi_i^{2r-3}$ is extremely small as compared to the other terms. In other words, even though the coupling may be relevant, we will restrict ourselves to integrating out modes with an IR cutoff such that quantum corrections to $h_i$ do not blow out of proportion and consequently change the nature of the fixed points. In the first order approximation, the scaling dimension of $h_i$ can thus be taken to be its classical mass dimension. 

Such a restriction of the coupling $h_i$ to simplify our calculation essentially means that we are looking at the phase transitions at a tunable mesoscopic limit, rather than in the exact thermodynamic limit (i.e. the IR cutoff $(l^*)$ can be taken large enough provided we appropriately fine-tune the contribution $h_i \phi_i^{2r-3}$). This assumption is reasonable because we want to capture details of higher order transitions very close to the critical point, i.e., $h_i \to 0$, and also, the physics of different phase transitions really depends on moving in the $t$ direction. Note that in the special case where $h_i =0$, the effective Hamiltonian in \eqref{LG1} for $r=3$ describes third order phase transitions in the exact thermodynamic limit, since $l^*$ can now go all the way up to $l^* \to \infty$, with the only exception being that critical exponents involving $h_i$ are not defined.

Here we will compute the critical exponents solely below four dimensions even though the upper critical dimension for $r=3$ is $d_u =6$. This is because the couplings $g_4$ and $g_6$ appearing in the phenomenological Hamiltonian for $r=3$ become irrelevant for $d>4$. Hence we cannot obtain critical exponents using linear stability analysis as done previously in \S \ref{r2d2} for $d>4$.

With these subtleties in mind, the stage is clear to calculate the critical exponents of $r=3$. We start with writing the beta functions of the couplings $g_4$ and $g_6$ which appear in \eqref{LG1} for $r =3$. The beta functions are given by:
\begin{eqnarray}
  \Lambda \frac{d g_{4}}{d \Lambda} &=& [d- 4] g_{4} + \frac{a(N+ 8)}{3}g_{4}^2 \label{bet3}\\
\Lambda \frac{d g_{6}}{d \Lambda} &=& [2d- 6] g_{6} + a(N+ 14) g_4g_6 - \frac{10 a (N+ 26)}{9}g_4^3 \label{bet31} 
 \end{eqnarray}
where $a$ is given in \eqref{a}. As in the case of $r =2$, we obtain two fixed points in this case as well. The trivial Gaussian fixed point is given by:
\beq
g_4^* = 0 \, , \qquad \qquad g_6^* = 0
\eeq
We now perform linear stability analysis about this fixed point by expanding the beta function near the same. We again look at RG flows in the vicinity of this fixed point just below four dimensions, where we have defined $\epsilon = 4- d$. We obtain the following linearized beta functions:
\beq
\begin{pmatrix}
   \beta(\delta g_4)\\
   \beta(\delta g_6)
\end{pmatrix} = \begin{pmatrix}
 -\epsilon \qquad &    0 \\
0 \qquad & 2(1- \epsilon) 
\end{pmatrix}\begin{pmatrix}
   \delta g_4\\
   \delta g_6
\end{pmatrix}
\eeq
Since $0< \epsilon \ll 1$, the coupling $g_4$ is relevant while $g_6$ is irrelevant near this fixed point. We use the relevant coupling to compute the critical exponents which are given below:
\beq
\begin{split}
    \nu = \frac{1}{\epsilon} \, &, \qquad \qquad \Delta =  \frac{1}{2} + \frac{1}{\epsilon}\\
    \alpha = 4 \lc 1-\frac{1}{\epsilon}\rc \, &, \qquad \qquad \beta = -\frac{1}{2} + \frac{1}{\epsilon} \\
     \gamma = 1 \, &, \qquad \qquad \delta = \frac{2+\epsilon}{2- \epsilon}
\end{split}
\eeq
Notice that for $d =3$ i.e. $\epsilon = 1$, the above values of critical exponents match the ones computed via saddle point approximation in Table \ref{table1}. This is because the one-loop corrections near the Gaussian fixed point do not change the scaling dimensions of the couplings. Hence we do not get any corrections to the saddle point result at this order. 

Notice that there is no $N$-dependence in the critical exponents near the Gaussian fixed point. This is because about the Gaussian fixed point, the terms in \eqref{bet3} and \eqref{bet31} proportional to $N$ are quadratic or higher order in $\delta g_4$ and $\delta g_6$. However since we are only considering linear terms in our analysis, these terms do not contribute and hence we do not get any $N$ dependence.

For $d<4$, there exists another non-trivial fixed point of the beta functions in \eqref{bet3}, which is given by: 
\beq
g_4^* = -\frac{3 (d-4)}{a (N+8)} \, , \qquad \qquad g_6^* = -\frac{30 (d-4)^3 (N+26)}{a^2 (N+8)^2 (6 (N+20) - d (N+26))}
\eeq
The fixed point takes the following form in $d = 4-\epsilon$ dimensions:
\beq
g_4^* = \frac{3 \epsilon}{a (N+8)} \, , \qquad \qquad g_6^* = \frac{15 \epsilon^3 (N+26)}{a^2 (N+8)^3}
\eeq
Linearizing the beta functions about this fixed point, we obtain:
\beq
\begin{pmatrix}
   \beta(\delta g_4)\\
   \beta(\delta g_6)
\end{pmatrix} = \begin{pmatrix}
 \epsilon \qquad &    0 \\
 O(\epsilon^2) \qquad & 2 + \epsilon \frac{N+26}{N+8}
\end{pmatrix}\begin{pmatrix}
   \delta g_4\\
   \delta g_6
\end{pmatrix}
\eeq
Notice that near this fixed point both the couplings are irrelevant. Hence the whole $g_4 -g_6$ plane is a basin of attraction. As we change the temperature i.e. vary the coupling $g_4$, the theory always flows to the above non-trivial fixed point. As stressed in \S\ref{scalerg}, we need at least one relevant coupling to write down the scaling relations, hence we cannot use the RG flow to obtain critical exponents near this fixed point. Hence the critical exponents for $r=3$ are determined by RG flows near the Gaussian fixed point.

\subsection{Critical exponents for fourth order phase transitions $(r=4)$}
We will now compute the critical exponents for fourth order phase transitions. The beta functions for the couplings $g_6$ and $g_8$ appearing in \eqref{LG1} for $r=4$ are given by:
\begin{eqnarray} \label{beta4}
\Lambda \frac{d g_{6}}{d \Lambda} &=& [2d- 6] g_{6} - \frac{a(N+ 6)}{7}g_8 \\
\Lambda \frac{d g_{8}}{d \Lambda} &=& [3d- 8] g_{8} + \frac{7a}{5}(N+ 24)g_6^2 
 \end{eqnarray}
 Again we have a trivial Gaussian fixed point, i.e.
  \beq
g_6^* = 0 \, , \qquad \qquad  g_8^* = 0\, .
 \eeq
 We perform the linear stability analysis about the Gaussian fixed point just below three dimensions, since by naive power-counting we know that these couplings are irrelevant for $d\geq 4$. Defining $\lambda \equiv 3 - d$, the linearized expansion of beta functions near the Gaussian fixed point is given by:
 \beq
\begin{pmatrix}
   \beta(\delta g_6)\\
   \beta(\delta g_8)
\end{pmatrix} = \begin{pmatrix}
 - 2\lambda \qquad &    -\frac{a}{7}(N+6)\\
0 \qquad & 1-3\lambda
\end{pmatrix}\begin{pmatrix}
   \delta g_6\\
   \delta g_8
\end{pmatrix}
\eeq
We see that the coupling $g_6$ is relevant while the coupling $g_8$ can either be relevant, irrelevant or marginal depending on $\lambda$. For small $\lambda$, only  $g_6$ is relevant, which we use to determine the critical exponents: 
\beq
\begin{split}
    \nu = \frac{1}{2\lambda} \, &, \qquad \qquad \Delta =  \frac{1 + 3\lambda}{4\lambda} \\
    \alpha = \frac{3}{2} \lc 3-\frac{1}{\lambda}\rc \, &, \qquad \qquad \beta = \frac{1 - \lambda}{4\lambda}\\
     \gamma = 1 \, &, \qquad \qquad \delta = \frac{1 + 3\lambda}{1-\lambda}
\end{split}
\eeq
The beta functions given by \eqref{beta4} also admit a non-trivial fixed point which is given by:
 \beq
g_6^* = -\frac{10 \left(3 d^2-17 d+24\right)}{a^2 \left(N^2+30 N+144\right)} \, , \qquad \qquad  g_8^* = -\frac{140 (d-3)^2 (3 d-8)}{a^3 (N+6)^2 (N+24)},
 \eeq
Notice that this fixed point exists only for $d<3$. For $d>3$, $g^*_8$ becomes negative, thereby making the Hamiltonian unbounded. Just below three dimensions, the fixed point is given by:
 \beq
 g_6^* = \frac{10 (1-3 \lambda ) \lambda }{a^2 \left(N^2+30 N+144\right)} \qquad \qquad g_8^* = \frac{140 \lambda ^2 (3 \lambda -1)}{a^3 (N+6)^2 (N+24)}
 \eeq
 Expanding the beta-function near the fixed point in $3- \lambda$ dimensions, we obtain the following linearized form:
 \beq
\begin{pmatrix}
   \beta(\delta g_6)\\
   \beta(\delta g_8)
\end{pmatrix} = \begin{pmatrix}
 -2\lambda \qquad &    -\frac{a}{7}(N+6)\\
\frac{28 \lambda}{a (N+6)} \qquad & 1-3\lambda
\end{pmatrix}\begin{pmatrix}
   \delta g_6\\
   \delta g_8
\end{pmatrix}
\eeq
The eigenvalues of above matrix are given by:
$$\frac{1}{2} \left(1 - 5\lambda -\sqrt{\lambda ^2-18 \lambda +1)}\right) \,  \qquad \text{and} \qquad \frac{1}{2} \left(1 -5 \lambda + \sqrt{\lambda ^2-18 \lambda +1}\right).$$
 Since $\lambda$ is small, we can linearly expand the above eigenvalues while ignoring higher order terms in $\lambda$. At the first order in $\lambda$, the eigenvalues are given by:
$$2\lambda + O(\lambda^2) \, , \qquad \qquad 1 - 2 \lambda + O(\lambda^2)$$
We find that both the eigenvalues are positive i.e. both the couplings $g_6$ and $g_8$ are irrelevant near the non-trivial fixed point. Hence the critical exponents are given by the values obtained near the Gaussian fixed point, similar to our discussion for $r=3$. Notice that just like the case for $r=3$, the critical exponents are independent of $N$ here as well.

For $r> 4$, the one-loop corrections to the beta functions of the corresponding couplings vanish. Hence we need to look at higher loop corrections to the beta function in order to obtain the values of critical exponents beyond saddle point. The main results of this section are summarized in Table \ref{table5}.
\begin{table}[h!]
\begin{center}
\begin{tabular}{|l| l| l| l| l| l| l|}
\hline
r & $\alpha$ & $\beta$ & $\gamma$ & $\delta$ & $\nu$  & $\Delta$ \\
  \hline
   &      &    & && &  \\
 2 & $\epsilon \lc\frac{1}{2} - \frac{N+2}{N+8}\rc $ &   $\frac{1}{2} - \frac{\epsilon}{4} \lc1 - \frac{N+2}{N+8}\rc $ & $1 + \frac{\epsilon}{2}\lc\frac{N+2}{N+8}\rc $ &$ 3 + \epsilon\lc \frac{N+2}{N+8}\rc $& $\frac{1}{2} + \frac{\epsilon}{4} \frac{ (N+2)}{(N+8)} $&$ \frac{3}{2} + \frac{ \epsilon}{4} \lc\frac{3(N+2)}{N+8} - 1 \rc $\\
  &      &    & && &  \\
 \hline
  &      &    & && &  \\
3 &$ 4 \lc 1-\frac{1}{\epsilon}\rc$ &$-\frac{1}{2} + \frac{1}{\epsilon}$&  1& $\frac{2+\epsilon}{2- \epsilon}$& $\frac{1}{\epsilon} $&$\frac{1}{2} + \frac{1}{\epsilon}$\\
&      &    & && &  \\
 &      &    & && &  \\
 \hline
  &      &    & && &  \\
4 &$\frac{3}{2} \lc 3-\frac{1}{\lambda}\rc$&$\frac{1 - \lambda}{4\lambda}$&1&$\frac{1 + 3\lambda}{1-\lambda}$& $\frac{1}{2\lambda}$& $\frac{1 + 3\lambda}{4\lambda} $\\
&      &    & && &  \\
\hline
\end{tabular}
\caption{One-loop corrections to the critical exponents. Here $\epsilon = 4-d$ and $\lambda = 3-d$. Note that the critical exponents for $r=2$ are calculated about the Wilson-Fisher fixed point, whereas the critical exponents for $r=3$ and $r=4$ are calculated about the Gaussian fixed point. This is due to the fact that for phase transitions of order $r=3,4$, there exists a relevant coupling only about the Gaussian fixed point.}
\label{table5}
\end{center}
\end{table}

 \subsection{Discussion of results for $r = 3,4$}
We now give a summary of results regarding critical exponents in various dimensions\footnote{We thank anonymous referee 2 for their insightful questions, which have been instrumental for the analysis in this subsection.}. As we pointed out in \S \ref{fluctuations}, saddle point calculations dominate over fluctuation contributions above $d=2r$. Below this upper critical dimension, saddle point dominates over fluctuations only if the generalized Ginzburg criteria is satisfied. If this is not the case, we need to take fluctuation contributions into account by venturing away from the saddle point, which gives rise to various scaling relations as derived in \S \ref{secscaling}. In order to calculate all possible critical exponents using these scaling relations, we need to estimate the values of two of them. In this section, we have evaluated the values of $\nu$ and $\Delta$, and consequently evaluate all of the critical exponents. 
 
In order to calculate all critical exponents using the scaling relations, we need to work within the regime where the coupling $t_r$ is relevant, which is always true for $r =2$, and conditionally true for $r=3$ and $4$ provided we work in $d \leq 4$ and $d \leq 3$ respectively. In this section, we have calculated $\nu$ and $\Delta$ using the one-loop beta functions for certain higher order phase transitions, i.e., for $r=2,3$ and $4$.
 
 We find that there exist two fixed points: one of them being the trivial Gaussian fixed point, i.e., the free theory limit, while the other point for $r=3,4$ below $d=4,3$ respectively is analogous to the Wilson-Fisher (WF) fixed point for $r=2$. We observe that within the $\epsilon$-expansion, there exist no relevant couplings about the WF fixed point for $r =3,4$ cases, in contrast to the $r =2$ case. Since there are no relevant directions, the scalings and scaling relations of physical observables derived in Section 4 are not valid near WF fixed point. Consequently we perform our analysis about the Gaussian fixed point for $r =3,4$. 

Note that $\epsilon$ is not the parameter characterizing our perturbative expansion for the $r=3,4$ cases as compared to the $\epsilon$-expansion performed for $r=2$. This is because the values of couplings at Gaussian fixed point are independent of $\epsilon$ (i.e., the couplings are zero), in contrast to the WF fixed point where the couplings depend on $\epsilon$. Consequently the critical exponents computed about the Gaussian fixed points are exact in $\epsilon$, in contrast to the computation of critical exponents in the expansion about WF point. Hence perturbation theory remains valid even when we take $\epsilon$ large enough, since expanding about the Gaussian fixed point is a perturbative expansion in the couplings, and not in terms of $\epsilon$ since the couplings are independent of $\epsilon$.
    
Our analysis about the Gaussian point for $r =3,4$ shows that some critical exponents diverge in the $\epsilon \, ( \text{or} \, \lambda) \to 0$ limit, as given in Table \ref{table5}. This is because the couplings become marginal when this limit is imposed, and consequently our scaling relations in \S \ref{secscaling} become ill-defined, since they are valid only under the assumption that the coupling $t_r$ is relevant. This suggests us to go to finite non-zero values of $\epsilon(\lambda)$ to ensure the relevance of couplings and the subsequent validity of scaling relations in \S \ref{secscaling}. The results given in Table \ref{table5} therefore make sense when $\epsilon (\lambda)$ is finitely non-zero, and not when $\epsilon (\lambda) \to 0$. Coincidentally, for $\epsilon = 1$, the one-loop results for $r=3$ match the expected saddle point results for third order phase transitions in $d=3$, and similarly for $\lambda = \frac{1}{3}$ for $r=4$. In short, the "divergence" of the critical exponents neither invalidates perturbation theory and nor the $\frac{1}{\epsilon}$ factor appearing in Table \ref{table5} suggests anything dramatic, but are essential features of the phase transition. Since perturbation theory remains valid, we strongly expect other perturbative methods, such as dimensional regularization to give us same results about the Gaussian fixed point. We also note that our results in this subsection are not entirely unexpected, given the classical mass dimensions of the couplings obtained in \eqref{couplingdimensions}.

Further, the absence of $N$-dependence in Table \ref{table5} is a standard feature of expansion about the Gaussian fixed point. It should be noted that the expansion about the Gaussian fixed point for $r=2$ does not lead to any $N$-dependence in the critical exponents as well. 

We also notice from Table \ref{table5} that if $\epsilon(\lambda)$ is close enough to zero, the critical exponent $\alpha$ can become drastically different from the saddle point expectation $\alpha =0$. In particular, an interesting case arises when $\abs{\alpha} \geq 1$, which basically implies that the order of the phase transition can potentially get changed under RG flows. Note that the changing of order of phase transition in this fashion is different from explicitly introducing lower order terms into \eqref{LG1}. 
   

\section{Summary and Discussion}
We briefly summarize our work and discuss our conclusions here. In our work, we look at higher order thermodynamic phase transitions. For phase transitions involving a local order parameter, we write a generalized phenomenological Landau Hamiltonian, which describes $r$th order phase transitions. Near criticality, these phase transitions are characterized by divergences in the physical observables. We capture such divergences by introducing critical exponents. As a first step, we calculate these critical exponents by using the saddle point approximation. 

Next, we investigate the role of fluctuations. We consider fluctuations giving rise to polynomials in the local field whose order is smaller than the order of terms in the Lagrangian. We show that such fluctuations lead to non-analyticity in the critical exponents. We also show that there is a generalization of the notion of upper and lower critical dimensions, which is a straightforward extension of the $r=2$ case. Above the upper critical dimension $d= 2r$, the saddle point calculation is valid since fluctuations do not introduce new singularities. Below the upper critical dimension, we generalize the Ginzburg criteria to precisely quantify when fluctuations dominate over the saddle point calculation.

Next, we introduce scaling forms for physical observables derived from the partition function. These scaling forms can be conveniently derived using Wilsonian renormalization. We use these forms to obtain scaling relations between the critical exponents, which continue to hold beyond the saddle point approximation. We show that given relevant couplings $t$ and $h$, one can use their mass dimensions and these scaling relations to determine all the critical exponents for these transitions. 

We further use the renormalization group to compute corrections to the critical exponents by calculating the one-loop beta functions. The one-loop beta functions are used to identify fixed points of the RG flow. We calculate the scaling dimensions of the couplings about these fixed points and use them to write the corrections to the critical exponents beyond the saddle point. For $r =3$ and $r=4$, we determine the corrected critical exponents of the couplings near the Gaussian fixed point. We also find a non-trivial fixed point in both cases. However, there are no relevant couplings in the vicinity of the non-trivial fixed point, and consequently, the sole corrections arise due to flows near the Gaussian fixed point. For $r \geq 5$, the one-loop beta function vanishes, and hence we need to go beyond one-loop calculations to find corrections to the critical exponents.

We now briefly discuss why we rarely observe higher order local phase transitions in physical systems. Higher order phase transitions with a local order parameter require a delicate fine-tuning which sets lower order terms in the Hamiltonian to zero. Such a fine-tuning is uncommon in most systems of physical interest, where such terms can arise due to fluctuations. These lower order terms can be avoided due to the existence of a symmetry, however we presently do not know any symmetry argument which explicitly rules out such terms.

A conceptual feature we wish to convey through our work is that the one-loop method used in our of calculation of critical exponents in \S \ref{crrg} is possibly more convenient than using Feynman diagrammatics to do the same, as performed for $r=2$ in \cite{Peskin:1995ev, kardar_2007}. One important question which remains to be answered is to estimate the contribution of higher loop corrections to the critical exponents.

We will list out possible avenues that branch out from our work. An immediate direction is to construct a guiding principle that classifies higher order non-local phase transitions. Using such a principle, we hope to write a phenomenological Hamiltonian describing higher-order non-local phase transitions. An important class of non-local third order phase transitions arise in  Gross-Witten-Wadia type models. In a forthcoming work, we will investigate such phase transitions and calculate critical exponents for the same. In general, one can investigate such phase transitions in the context of finite-range Coulomb gas models. We visualize our present work to be a starting basis for looking at these problems and hope to uncover some of the above mysteries in future works. 

\section*{Acknowledgements}
We thank Abhishek Dhar and Akhil Sivakumar, and especially anonymous referee for their comments on the draft. We are also grateful to Spenta Wadia for going through the earlier version of the draft meticulously and for suggesting corrections. The authors acknowledge gratitude to the people of India for their steady and generous support to research in basic sciences.

 \appendix

\section{Why critical exponents are single valued in the $(h,t)$ plane?}
\label{exponentspm}
In this appendix, we will see that single valuedness of critical exponents in the $(h,t)$ plane follows from the analyticity assumption combined with our scaling conjecture. Let us consider a more general scaling form such that $C^{\pm}_V, \alpha_{\pm}, g_{\pm}, \Delta_{\pm}$ are different for $t>0$ and $t<0$ respectively. As an example, we will consider that the singular part of specific heat is given by the following function
\beq
C^{\pm}_V \sim \abs{t}^{r-\alpha_{\pm} -2} \widehat{g}_{\pm}\lc \frac{h}{t^{\Delta_{\pm}}}\rc.
\eeq
However we argue that this is ruled out by free energy being analytic everywhere apart from $h=0$ and $t<0$. To understand this, consider a point in the $(h,t)$ plane at $t=0$ and finite $h$. Since the function $C$ is analytic in the vicinity of this point, it can be expanded in terms of a Taylor series
\beq
C_V(t\ll h^{\Delta}) = A(h) + t B(h) + \text{O}(t^2) + \dots \label{cvtaylor}
\eeq
Since we already have a power series description, $C$ can be obtained from the two power series expansions given by
\beq
C_V^{\pm} = \abs{t}^{r-\alpha_{\pm}-2} \ls A_{\pm} \lc\frac{h}{t^{\Delta_{\pm}}}\rc^{p_{\pm}} + B_{\pm} \lc\frac{h}{t^{\Delta_{\pm}}}\rc^{q_{\pm}} + \dots \rs,
\eeq
where $\lc p_{\pm}, q_{\pm}\rc$ are powers of the largest terms of $\widehat{g}_{\pm}$. We will now match the expressions for $C_V^{\pm}$ with the power series given in \eqref{cvtaylor}, which leads to the identification
\beq
r-\alpha_{\pm}-2 = - p_{\pm} \Delta_{\pm} \quad \& \quad r-\alpha_{\pm}-1 = -q_{\pm} \Delta_{\pm}
\eeq
Thus the series expansions can be written as
\beq
C_V^{\pm} (t\ll h^{\Delta}) = A_{\pm}\, h^{\frac{-(r-\alpha_{\pm}-2)}{\Delta_{\pm}}} + B_{\pm}\, \abs{t}\, h^{\frac{-(r-\alpha_{\pm}-1)}{\Delta_{\pm}}} + \dots
\eeq
As a consequence of continuity at $t=0$, we can use the equation above to obtain the following relations between the critical exponents
\beq
\alpha_{+} = \alpha_{-} \equiv \alpha, \quad \Delta_+ = \Delta_- \equiv \Delta,
\eeq
and so on for other critical exponents as well since these arguments generalize for any scaling function $X(t,h)$ which possess analyticity everywhere apart from the line $h=0$ and $t<0$.

\section{Kinetic fluctuations in $d\leq 2$ and loss of order}\label{kf}

Consider the Hamiltonian \eqref{LG1} where the order parameter given by $\vec{\phi} = \phi\hat{\phi}$ is $N$-dimensional, i.e. the target space is $N$-dimensional. Taking the saddle point of the partition function fixes the magnitude of the order parameter, thereby spontaneously breaking the O($N$) symmetry to O$(N-1)$, with the $N-1$ angles remain unconstrained. Denoting these angles by $\Theta^{\alpha}$ and the metric on the target space as $g_{\alpha \beta}$, we can rewrite the kinetic part of the Hamiltonian, which describes the Goldstone modes parametrized by $\Theta^{\alpha}$ as

\begin{equation}
 \beta H = \int d^dx \left[\frac{K}{2} \phi^2 g_{\alpha \beta} \, \frac{\partial \Theta^{\alpha}}{\partial x^{i}} \frac{\partial \Theta^{\beta}}{\partial x^{i}} +   t_r \, \phi^{2(r-1)}(x) + u_r \, \phi^{2r}(x)   - (h.\phi) \, \phi^{2(r-2)}\right].
\end{equation}

Our analysis can be made easier by considering physical target spaces where $g_{\alpha \beta}$ has no off-diagonal terms. In this case, the two point correlator can be found out by solving the Green function equation and is given by
\beq
\braket{\Theta^{\alpha}(x) \Theta^{\beta}(x')} = -\frac{2C(x-x')}{K\phi^2 \, g_{\alpha \beta}},
\eeq
 where $C(x)$ in the long distance limit is given by 
 \beq \label{Goldstone}
 \lim_{x \to \infty} C(x) = \begin{cases} 0 \quad &d>2\\
 \dfrac{x^{2-d}}{(2-d)S_d} \quad &d<2\\
 \dfrac{\ln{x}}{2\pi} \quad &d=2\\
 \end{cases}
 \eeq
Thus we see that kinetic fluctuations or Goldstone modes destroy long range order in $d\leq 2$, as correlators in these dimensions grow with distance. This leads to a destruction of order in $d\leq 2$ for systems possessing continuous symmetry, which defines the lower critical dimension. The analysis in this subsection is essentially the content of the Mermin Wagner theorem \cite{PhysRevLett.17.1133, Coleman:1973ci}, since Goldstone modes in the system would have infrared divergent two-point correlator as given by \eqref{Goldstone}. For discrete systems, the lower critical dimension is $d_l =1$, the corresponding argument remains unchanged as well for the case of higher order phase transitions.

\bibliographystyle{JHEP}
\bibliography{citation.bib}
\end{document}